\documentclass[12pt,preprint]{elsarticle}
\usepackage[latin9]{inputenc}
\usepackage{amsmath}
\usepackage{amssymb}
\usepackage{graphicx}

\makeatletter

\providecommand{\tabularnewline}{\\}

\newcounter{bla}

\journal{Computer Physics Communications}

\makeatother

\begin{document}
\begin{frontmatter}



\title{JefiGPU: Jefimenko's Equations on GPU}


\author[b]{Jun-Jie Zhang\corref{author}}

\author[b]{Jian-Nan Chen}

\author[a,b]{Guo-Liang Peng}

\author[b]{Tai-Jiao Du}

\author[b]{Hai-Yan Xie}

\cortext[author]{Corresponding author.\\
\textit{E-mail address:} zjacob@mail.ustc.edu.cn}

\address[a]{Beijing Institute of Technology, Beijing 100081, China}
\address[b]{Northwest Institute of Nuclear Technology, Xi'an 710024, China}


\begin{abstract}
We have implemented a GPU version of the Jefimenko's equations ---
JefiGPU. Given the proper distributions of the source terms $\rho$
(charge density) and $\mathbf{J}$ (current density) in the source
volume, the algorithm gives the electromagnetic fields in the observational
region (not necessarily overlaps the vicinity of the sources). To
verify the accuracy of the GPU implementation, we have compared the
obtained results with that of the theoretical ones. Our results show
that the deviations of the GPU results from the theoretical ones are
around 5\%. Meanwhile, we have also compared the performance of the
GPU implementation with a CPU version. The simulation results indicate
that the GPU code is significantly faster than the CPU version. Finally,
we have studied the parameter dependence of the execution time and
memory consumption on one NVIDIA Tesla V100 card. Our code can be
consistently coupled to RBG (Relativistic Boltzmann equations on GPUs)
and many other GPU-based algorithms in physics.
\end{abstract}

\end{frontmatter}

% \linenumbers

% All CPiP articles must contain the following
% PROGRAM SUMMARY.

\textbf{PROGRAM SUMMARY/NEW VERSION PROGRAM SUMMARY} %Delete as appropriate.

\begin{small} {\em Program Title:} JefiGPU\\
 {\em CPC Library link to program files:} (to be added by Technical
Editor) \\
 {\em Developer's repository link:} (if available) \\
 {\em Code Ocean capsule:} (to be added by Technical Editor)\\
 {\em Licensing provisions(please choose one):} Apache-2.0 \\
 {\em Programming language:} Python\\
 {\em Supplementary material:} \\
 % Fill in if necessary, otherwise leave out.
{\em Journal reference of previous version:}{*} \\
 %Only required for a New Version summary, otherwise leave out.
{\em Does the new version supersede the previous version?:}{*}
\\
 %Only required for a New Version summary, otherwise leave out.
{\em Reasons for the new version:{*}}\\
 %Only required for a New Version summary, otherwise leave out.
{\em Summary of revisions:}{*}\\
 %Only required for a New Version summary, otherwise leave out.
{\em Nature of problem(approx. 50-250 words):}Jefimenko's equations
are numerically stable with given sources $\rho$ (charge density)
and $\mathbf{J}$ (current density). Providing proper sources, Jefimenko's
equations can give the electromagnetic fields without boundary conditions.
However, the relevant integrations of the Jefimenko's equations involve
a 4D space-time which is both memory and time consuming. With the
help of the state-of-art GPU technique, these integrations can be
evaluated with acceptable accuracy and execution time. In the present
work, we have implemented a GPU version of the Jefimenko's equations
and tested the accuracy and the performance of the code. Our work
have demonstrated a significant improvement of the performance compared
with a similar CPU implementation.\\
 %Describe the nature of the problem here. \\
{\em Solution method(approx. 50-250 words):}The GPU code mainly
deals with the implementation of the integrations in Jefimenko's equations.
Each CUDA kernel corresponds to a spatial point of $\mathbf{E}$ and
$\mathbf{B}$. The integrations involving the source region and the
retarded time are performed in each CUDA kernel. Ray, Numba and Cupy
packages are used to manipulate the scaling of GPU allocation and
evaluation.\\
 %Describe the method solution here.
{\em Additional comments including restrictions and unusual features
(approx. 50-250 words):}\\
 %Provide any additional comments here.
 \\

\noindent \\
 \end{small}


\section{Introduction}

Electromagnetic force, as one of the four basic interactions, plays
an important role in almost all fields in nature. The governing equations
of the classical electromagnetic force are the well-known Maxwell's
equations. To solve the Maxwell's equations numerically, various algorithms
including open-source\cite{Warren2019,Warren2016,Fedeli2019} and
commercial software\cite{Yoon2021,EMWorks,Lopez2010,remcom} have
been designed for both general and specific usages. Among the many
approaches to solve the Maxwell's equations, the Jefimenko's equations\cite{Jefimenko1989,Griffiths1999},
known as the general solutions of the Maxwell's equations, are the
ones that require only the spatial distribution of the source terms
$\rho$ (charge density) and $\mathbf{J}$ (current density) to give
the relevant electromagnetic (EM) fields. 

The benefits of the Jefimenko's equations are two folds. First, they
contain two integrals of $\rho$ and $\mathbf{J}$. Combined with
the continuum relation $\nabla\cdot\mathbf{J}+\partial\rho/\partial t=0$,
these integrals are equivalent to the Maxwell's equations. Integrals
are much more stable than differences. Compared with the FTDT (Finite-Difference
Time-Domain)\cite{Yee1966,Oskooi2010,Warren2019} and FDFD (Finite-Difference
Frequency- Domain)\cite{Champagne2001} methods, Jefimenko's equations
do not have the trouble of numerical blow up (this is only true when
we use the concept of charge screening\cite{Peskin2018}). Second,
the Jefimenko's equations do not need the boundary conditions ---
as long as one provides the proper distributions of the sources. The
$\mathbf{E}$ and $\mathbf{B}$ fields can be obtained by performing
an integration in the vicinity of the sources. This is convenient
especially when we calculate the EM fields in a fixed box, since in
FEM (Finite Element Method)\cite{Otin2015,Jin2014,Delisle1991}, FDTD
and FDFD one has to carefully design the boundary conditions\cite{Komatitsch2003,Tang2012,Engquist1977}
to absorb the reflections of the EM fields. 

However, the Jefimenko's equations also encounter certain difficulties.
The integral is performed over the entire source volume space with
retarded time. This requires all the information of the sources in
the relevant space-time, and is therefore memory demanding. Meanwhile,
the denominator of the integrated involves a divergence at $\mathbf{r}=\mathbf{r}^{\prime}$
(see Esq. (\ref{eq:E}) \textasciitilde{} \ref{eq:tr}). The divergence
is somewhat troublesome in numerical evaluations. Thus, the refinement's
equations are usually used to solve far-field EM fields\cite{Baev2019,Kuznetsov2018}.
Except for the numerical difficulties, the source term $\rho$ may
not be easily available since it is experimentally hard to measure
the charge density distributions\cite{Shao2016}. To provide a distribution
of the charge and current densities, one may need to solve the transport
equations (e.g., the Boltzmann equation \cite{Zhang2020}) which is
another challenge so far.

In this regard, we have applied the Jefimenko's equation on the state-of-art
GPU based on our previous work ZMCintegral\cite{Wu2019tsf,zhang2020tsf}
and RBG\cite{Zhang2020}. Our present code mainly contributes to the
field of Heavy-Ion collisions\cite{Yagi2008,Andersson2007,Zhao2020,Fukushima2016}
where the electromagnetic fields in the early time significantly influence
the evolution of the quark-gluon matter. Our hope is to combine the
Jefimenko's equations to the relativistic Boltzmann equation\cite{Zhang2020}
and eventually answer the question of ``early thermalization''\cite{Berges:2020fwq,Ryblewski2010,Heinz2002}.
Due to the high clock rate, high instruction per cycle\cite{Gorelick2014},
and multiple cores of the GPUs, more and more numerical algorithms
that were hard in the past are now possible\cite{Wu2019tsf,Zhang2019a}.
In the present work, we use the Python package Ray\cite{PMoritz2018},
Numba\cite{Lam2015} and Cupy\cite{cupy_learningsys2017} to perform
the jefimenko's equations on GPU. Our work has demonstrated significant
performance improvement compared with the usual CPU implementation.
The results are stable and within an acceptable margin of error. 

The paper is organized in the following structure. In Sec. \ref{sec:GPU-Implementation},
we introduce the expressions of the Jefimenko's equations and the
details of the GPU implementation. In Sec. \ref{sec:Comparisons-with-CPU},
we compare the GPU results with the theoretical results. The corresponding
execution time of the GPU code is also compared with a similar CPU
implementation. In Sec. \ref{sec:Parameter-dependence-of}, we have
investigated the parameter dependence of the GPU code on one Tesla
V100 card, and presented the time consumption of HToD (Host To Device),
DToH (Device To Host) and the evaluation. The conclusion and outlook
of JefiGPU is in Sec. \ref{sec:Conclusion}. Through out the paper
we use the Rationalized-Lorentz-Heaviside-QCD unit.

\section{GPU Implementation\label{sec:GPU-Implementation}}

\subsection{Jefimenko's equations}

The Jefimenko's equation can be directly derived from the retarded
potential solution of Maxwell's equations\cite{Shao2016}. In Rationalized-Lorentz-Heaviside-QCD
unit, the Jefimenko's equations take the following form
\begin{eqnarray}
\mathbf{E}(\mathbf{r},t) & = & \frac{1}{4\pi}\int\left[\frac{\mathbf{r}-\mathbf{r}^{\prime}}{|\mathbf{r}-\mathbf{r}^{\prime}|^{3}}\rho(\mathbf{r}^{\prime},t_{r})\right.\nonumber \\
 &  & +\frac{\mathbf{r}-\mathbf{r}^{\prime}}{|\mathbf{r}-\mathbf{r}^{\prime}|^{2}}\frac{\partial\rho(\mathbf{r}^{\prime},t_{r})}{\partial t}\nonumber \\
 &  & \left.-\frac{1}{|\mathbf{r}-\mathbf{r}^{\prime}|}\frac{\partial\mathbf{J}(\mathbf{r}^{\prime},t_{r})}{\partial t}\right]d^{3}\mathbf{r}^{\prime}\label{eq:E}\\
\mathbf{B}(\mathbf{r},t) & = & -\frac{1}{4\pi}\int\left[\frac{\mathbf{r}-\mathbf{r}^{\prime}}{|\mathbf{r}-\mathbf{r}^{\prime}|^{3}}\times\mathbf{J}(\mathbf{r}^{\prime},t_{r})\right.\nonumber \\
 &  & +\left.\frac{\mathbf{r}-\mathbf{r}^{\prime}}{|\mathbf{r}-\mathbf{r}^{\prime}|^{2}}\times\frac{\partial\mathbf{J}(\mathbf{r}^{\prime},t_{r})}{\partial t}\right]d^{3}\mathbf{r}^{\prime}\label{eq:B}\\
t_{r} & = & t-|\mathbf{r}-\mathbf{r}^{\prime}|,\label{eq:tr}
\end{eqnarray}
where $\mathbf{E}$ and $\mathbf{B}$ are the electric and magnetic
field at space-time point $(\mathbf{r},t)$, $\rho$ and $\mathbf{J}$
are the charge and current densities, and $t_{r}$ is the retarded
time. Once the space-time distributions of $\rho$ and $\mathbf{J}$
are given, one can use Eqs. (\ref{eq:E}) \textasciitilde{} \ref{eq:tr}
to obtain the field distributions in space-time. 

The numerical difficulties of the Jefimenko's equations lie in the
following two aspects. First, to obtain the numerical value of $\mathbf{E}$
and $\mathbf{B}$ at each point in the observational region (not necessarily
overlaps the source region), an integration involving a 4-Dimensional
(4D) space-time is required. Supposing the observational region has
$n_{x,\text{o}}\times n_{y,\text{o}}\times n_{z,\text{o}}=100^{3}\sim10^{6}$
points, we need to perform $10^{6}$ integrations at each time step.
If each integration involves a summation of $n_{x,\text{s}}\times n_{y,\text{s}}\times n_{z,\text{s}}=100^{3}\sim10^{6}$
points, the task will be both time and memory consuming. Second, Jefimenko's
equations encounter a divergence at $\mathbf{r}=\mathbf{r}^{\prime}$
when the observational region overlaps the source region. The near-source
divergence has its profound physical origins and the usual treatment
is to adopt the concept of coarse screening\cite{Peskin2018}, which
is constrained by the grid size. With coarse screening, the $\mathbf{E}$
and $\mathbf{B}$ fields in a specific grid can only be generated
by the charge and density sources from other grids. If one is working
in the far-field calculations, then the divergence will not be a problem.

\subsection{GPU implementation of Jefimenko's equations}

We define two spatial regions of sizes $n_{x,\text{s}}\times n_{y,\text{s}}\times n_{z,\text{s}}$
and $n_{x,\text{o}}\times n_{y,\text{o}}\times n_{z,\text{o}}$, where
the subscripts $\text{o}$ and $\text{s}$ denote the physical quantities
in the source and observational regions, receptively. The infinitesimal
differences in these two regions are denoted as $dx_{\text{s}}\times dy_{\text{s}}\times dz_{\text{s}}$
and $dx_{\text{o}}\times dy_{\text{o}}\times dz_{\text{o}}$. In the
current work, we only use one NVIDIA V100 card to test the code, and
the functionality of multi-GPU on clusters is also supported via the
Ray package.

For numerical convenience, we discretize Eqs. (\ref{eq:E}) \textasciitilde{}
\ref{eq:tr} as
\begin{eqnarray}
\mathbf{E}(\mathbf{r},t) & = & \frac{1}{4\pi}\Sigma_{i,j,k}\left[\frac{\mathbf{r}-\mathbf{r}_{i,j,k}^{\prime}}{|\mathbf{r}-\mathbf{r}_{i,j,k}^{\prime}|^{3}}\rho(\mathbf{r}_{i,j,k}^{\prime},t_{r})\right.\nonumber \\
 &  & +\frac{\mathbf{r}-\mathbf{r}_{i,j,k}^{\prime}}{|\mathbf{r}-\mathbf{r}_{i,j,k}^{\prime}|^{2}}\frac{\rho(\mathbf{r}_{i,j,k}^{\prime},t_{r})-\rho(\mathbf{r}_{i,j,k}^{\prime},t_{r}-dt)}{dt}\nonumber \\
 &  & \left.-\frac{1}{|\mathbf{r}-\mathbf{r}_{i,j,k}^{\prime}|}\frac{\mathbf{J}(\mathbf{r}_{i,j,k}^{\prime},t_{r})-\mathbf{J}(\mathbf{r}_{i,j,k}^{\prime},t_{r}-dt)}{dt}\right]d\Omega^{\prime}\label{eq:E-1}\\
\mathbf{B}(\mathbf{r},t) & = & -\frac{1}{4\pi}\Sigma_{i,j,k}\left[\frac{\mathbf{r}-\mathbf{r}_{i,j,k}^{\prime}}{|\mathbf{r}-\mathbf{r}_{i,j,k}^{\prime}|^{3}}\times\mathbf{J}(\mathbf{r}_{i,j,k}^{\prime},t_{r})\right.\nonumber \\
 &  & +\left.\frac{\mathbf{r}-\mathbf{r}_{i,j,k}^{\prime}}{|\mathbf{r}-\mathbf{r}_{i,j,k}^{\prime}|^{2}}\times\frac{\mathbf{J}(\mathbf{r}_{i,j,k}^{\prime},t_{r})-\mathbf{J}(\mathbf{r}_{i,j,k}^{\prime},t_{r}-dt)}{dt}\right]d\Omega^{\prime}\label{eq:B-1}\\
t_{r} & = & t-|\mathbf{r}-\mathbf{r}_{i,j,k}^{\prime}|,\label{eq:tr-1}
\end{eqnarray}
where index $i\in\{1,2,...,n_{x}\}$, $j\in\{1,2,...,n_{y}\}$, $k\in\{1,2,...,n_{z}\}$
and volume element $d\Omega^{\prime}=dx^{\prime}dy^{\prime}dz^{\prime}$. 

\begin{figure}
\begin{centering}
\includegraphics[scale=0.5]{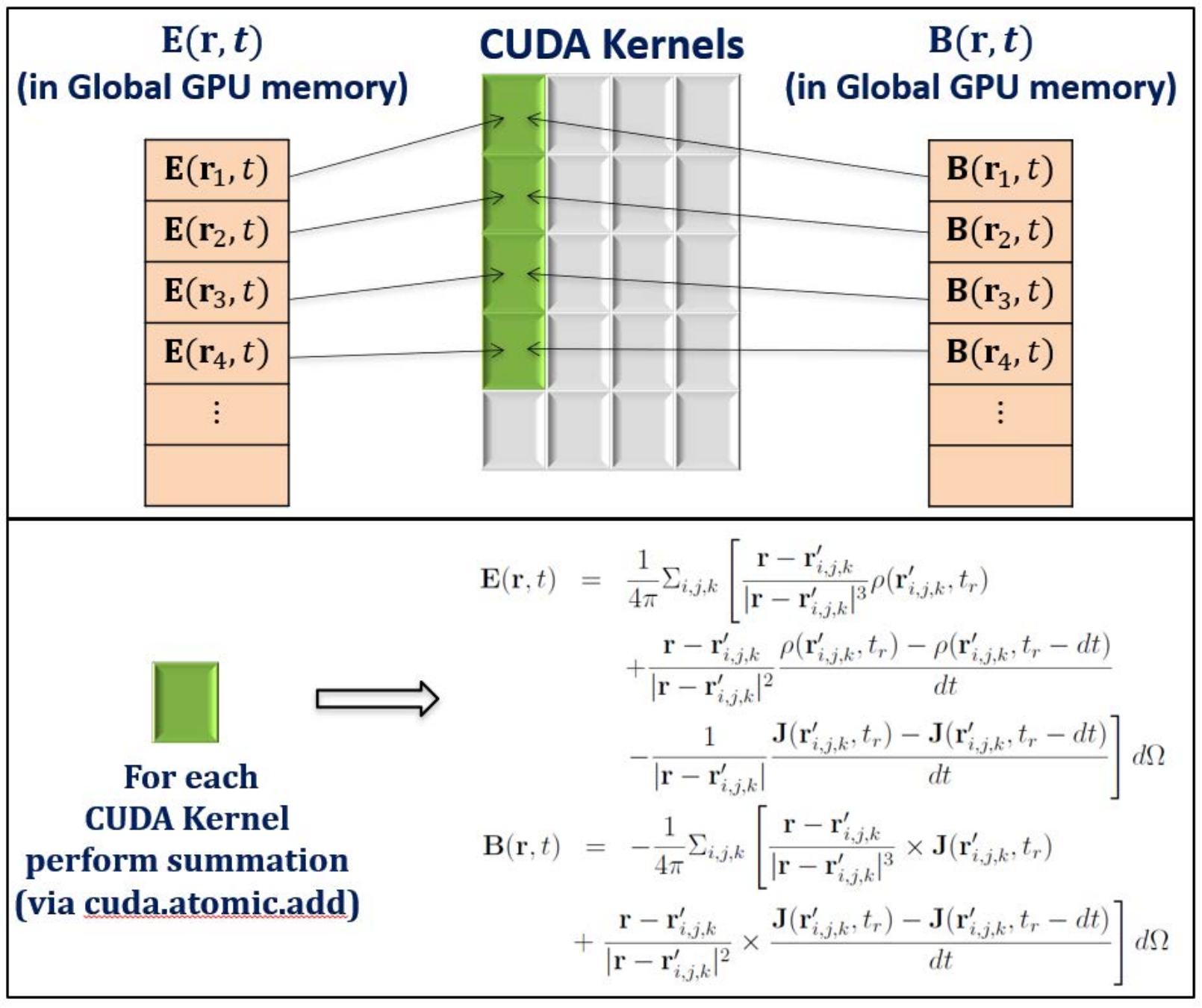}
\par\end{centering}
\caption{Schematic of the GPU implementation of Jefimenko's equations. Each
CUDA kernel corresponds to the evaluations of $\mathbf{E}(\mathbf{r}_{i},t)$
and $\mathbf{B}(\mathbf{r}_{i},t)$ at space-time point $(\mathbf{r}_{i},t)$
where $i$ is the index of the spatial grid. The integration at point
$(\mathbf{r}_{i},t)$ is performed in each CUDA kernel via Eqs. (\ref{eq:E-1})
\textasciitilde{} (\ref{eq:tr-1}).\label{fig:Schematic-of-the}}
\end{figure}

The single GPU implementation is illustrated in Fig. \ref{fig:Schematic-of-the}.
In each CUDA kernel we perform an integration to obtain $\mathbf{E}(\mathbf{r}_{i},t)$
and $\mathbf{B}(\mathbf{r}_{i},t)$ at space-time point $(\mathbf{r}_{i},t)$
where $i$ is the index of the spatial grid. At each call of the integration
method (jargon in Python) we claim an empty space for $\mathbf{E}(\mathbf{r},t)$
and $\mathbf{B}(\mathbf{r},t)$ in the global GPU memory. The calculated
EM fields via Eqs (\ref{eq:E-1}) \textasciitilde{} (\ref{eq:tr-1})
will be stored in this empty space. After the evaluation of the integral
(i.e., summation of Eqs. (\ref{eq:E-1}) \textasciitilde{} (\ref{eq:tr-1})),
new values of $\mathbf{E}(\mathbf{r},t)$ and $\mathbf{B}(\mathbf{r},t)$
at time snapshot $t$ will be obtained (i.e., stored in the pre-allocated
empty space) and transferred to the host memory. Since $\mathbf{E}(\mathbf{r},t)$
and $\mathbf{B}(\mathbf{r},t)$ are frequently transferred to the
host at each time, we store them in the global GPU memory for fast
data transferring. The numerical integration of Eqs. (\ref{eq:E-1})
\textasciitilde{} (\ref{eq:tr-1}) in each CUDA kernel requires a
full access of $\rho(\mathbf{r}^{\prime},t_{r})$ and $\mathbf{J}(\mathbf{r}^{\prime},t_{r})$
at all relevant space-time points (see Sec. \ref{sec:Parameter-dependence-of}
for details), thus $\rho$ and $\mathbf{J}$ should be stored in the
global GPU memory. Meanwhile, the GPU memory corresponds to $\rho$
and $\mathbf{J}$ will not be released after each call of the integration
method. Other temporary variables such as $\mathbf{r}^{\prime}$ and
$t_{r}$ are defined in local GPU memories. 

\begin{figure}
\begin{centering}
\includegraphics[scale=0.6]{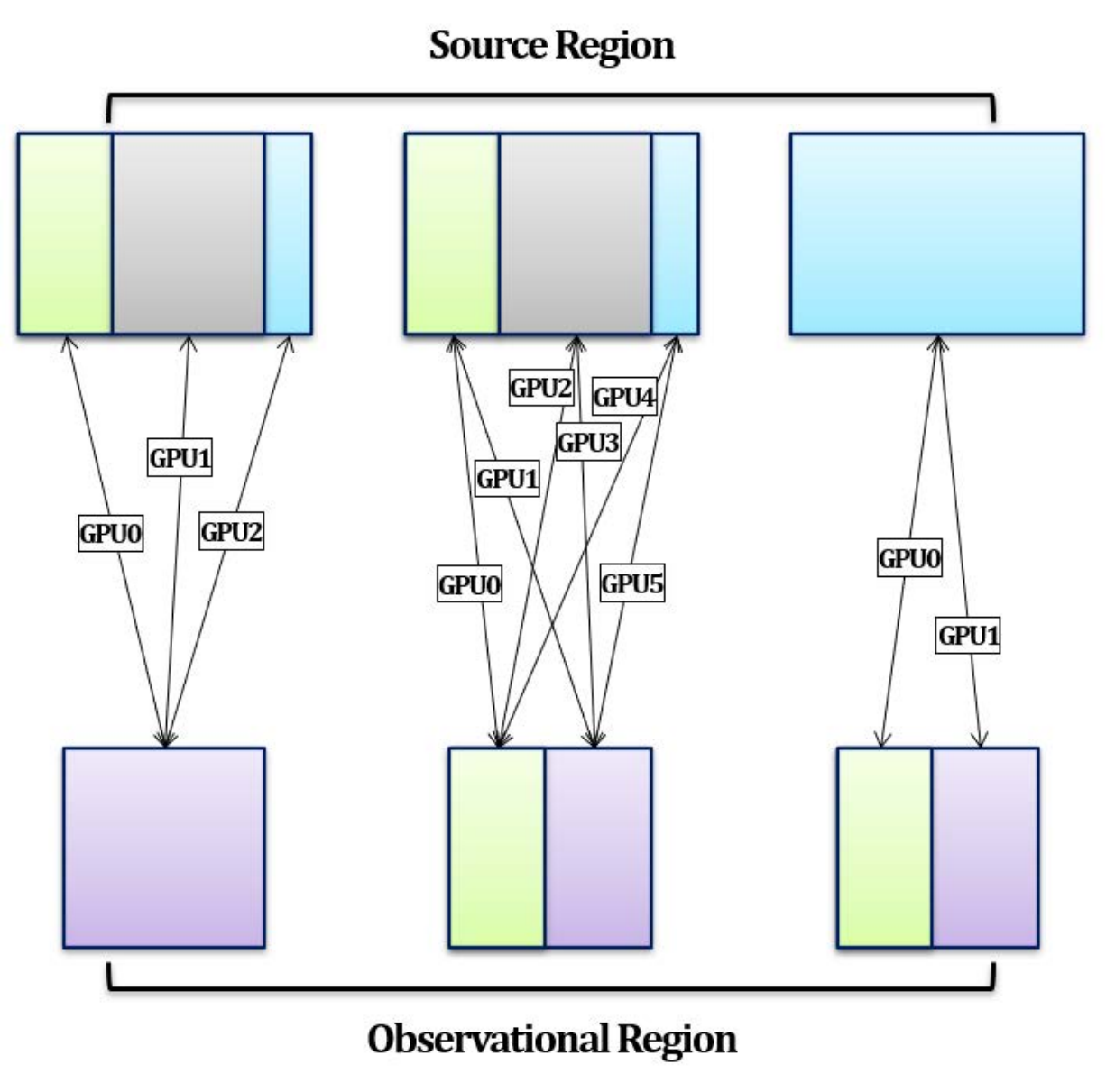}
\par\end{centering}
\caption{Three approaches for GPU scaling. The source and observational regions
are divided into several sub-regions. The sub-regions will form many
S-O pairs. Each S-O pair can be performed on one GPU card using the
algorithm illustrated in Fig. \ref{fig:Schematic-of-the}.\label{fig:Three-approaches-for}}

\end{figure}

The scaling of JefiGPU on GPU clusters is straight forward --- the
Ray package provides a succinct API to manipulate multiple GPU cards
and cluster nodes. There are three approaches to conduct the GPU scaling,
as is illustrated in Fig. \ref{fig:Three-approaches-for}. We can
divide the regions into several rectangular regions of S-O (Source
region and Observational region) pairs. Each pair can be performed
on one GPU card. In the code, we provide the users with a Python API
to manipulate these procedures.

\section{Comparisons with the theoretical results and the CPU based algorithm\label{sec:Comparisons-with-CPU}}

To see the performance of the GPU based algorithm, we will compare
it with the corresponding CPU implementation. The hardware set-up
of the two cases are summarized in Tab. \ref{tab:Hardware-set-up-of}.
The CPU code is performed via Python with the similar algorithm stated
in Fig. \ref{fig:Schematic-of-the}. The theoretical results are obtained
in Mathematica with the built-in function NIntegrate using the GlobalAdaptive
method. To avoid near-source divergence in the numerical integrations
of Eqs. (\ref{eq:E}) \textasciitilde{} \ref{eq:tr} in NIntegrate,
we introduce a screening volume of size $\delta^{3}$. Thus both $\rho$
and $\mathbf{J}$ give zero contribution to $\mathbf{E}$ and $\mathbf{B}$
when $|i-i^{\prime}|\leq\delta$, where $i\in\{x,y,z\}$. 
\begin{table}
\caption{Hardware set-up of the two conditions.\label{tab:Hardware-set-up-of}}

\centering{}%
\begin{tabular}{|c|c|c|}
\hline 
 & Hardware of CPU & Hardware of GPU\tabularnewline
\hline 
\hline 
GPU & Intel(R) Xeon(R) Silver 4110 & NVIDIA Tesla V100\tabularnewline
Implementation & CPU@2.10 GHz with 32 cores & with 32Gb memory\tabularnewline
\hline 
CPU & Intel(R) Xeon(R) Silver 4110 & None\tabularnewline
Implementation & CPU@2.10 GHz with 1 core & \tabularnewline
\hline 
\end{tabular}
\end{table}

\subsection{Constant sources\label{subsec:Constant-sources}}

We will firstly use the constant sources to see whether the GPU-Code
gives stable and acceptable results, and then compare the execution
time of the GPU-Code to the CPU-Code. The sources of $\rho$ and $\mathbf{J}$
take the following form
\begin{eqnarray}
\rho(x,y,z,t) & = & \begin{cases}
0 & \text{if }t<0\\
1 & \text{if }t\geq0
\end{cases}\nonumber \\
\mathbf{J}(x,y,z,t) & = & \begin{cases}
\mathbf{0} & \text{if }t<0\\
\mathbf{1} & \text{if }t\geq0
\end{cases},\label{eq:const_sources}
\end{eqnarray}
where $\mathbf{0}=(0,0,0)$ and $\mathbf{1}=(1,1,1)$. 

\begin{figure}
\begin{centering}
\begin{tabular}{ccc}
\includegraphics[scale=0.25]{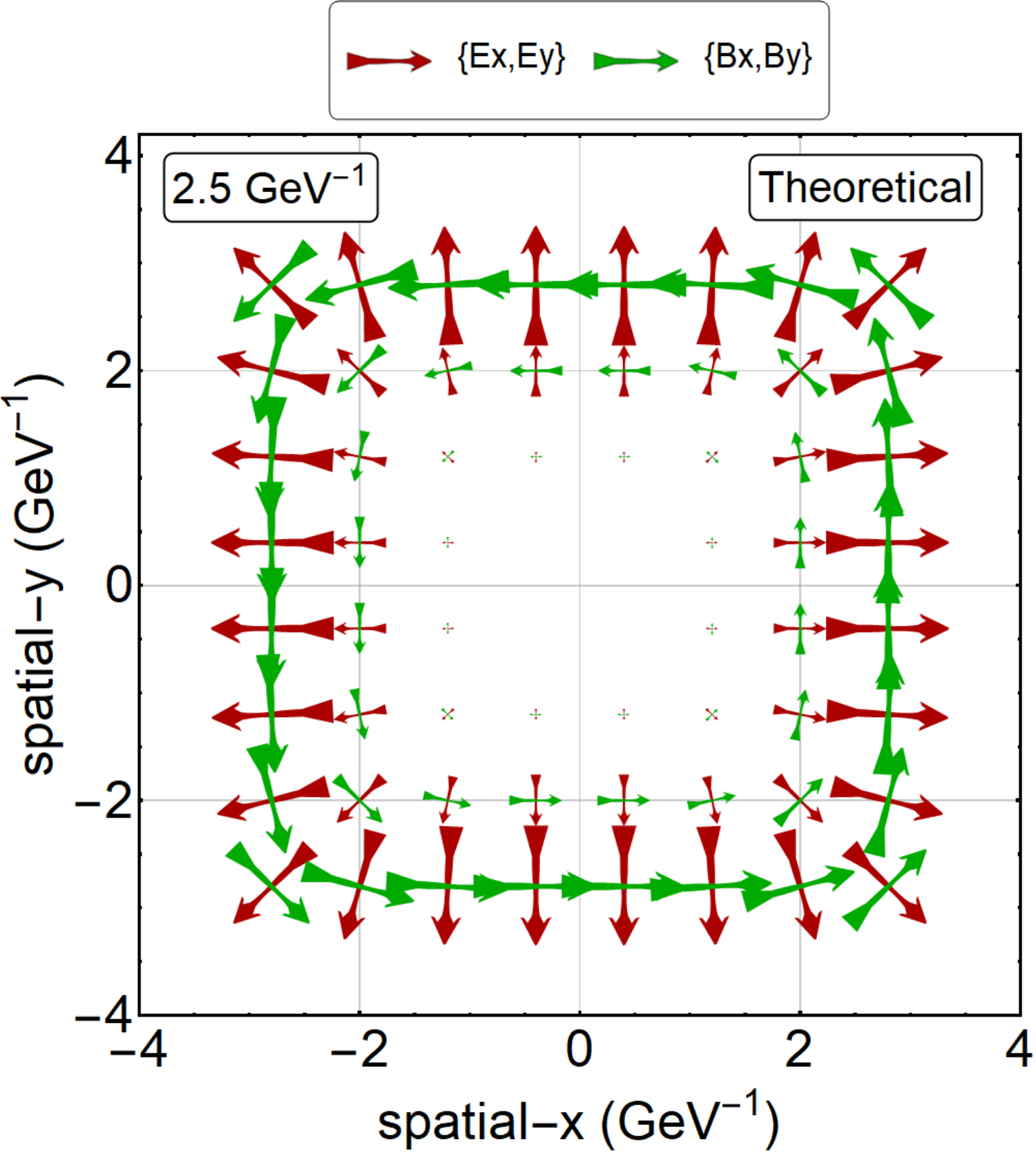} & \includegraphics[scale=0.25]{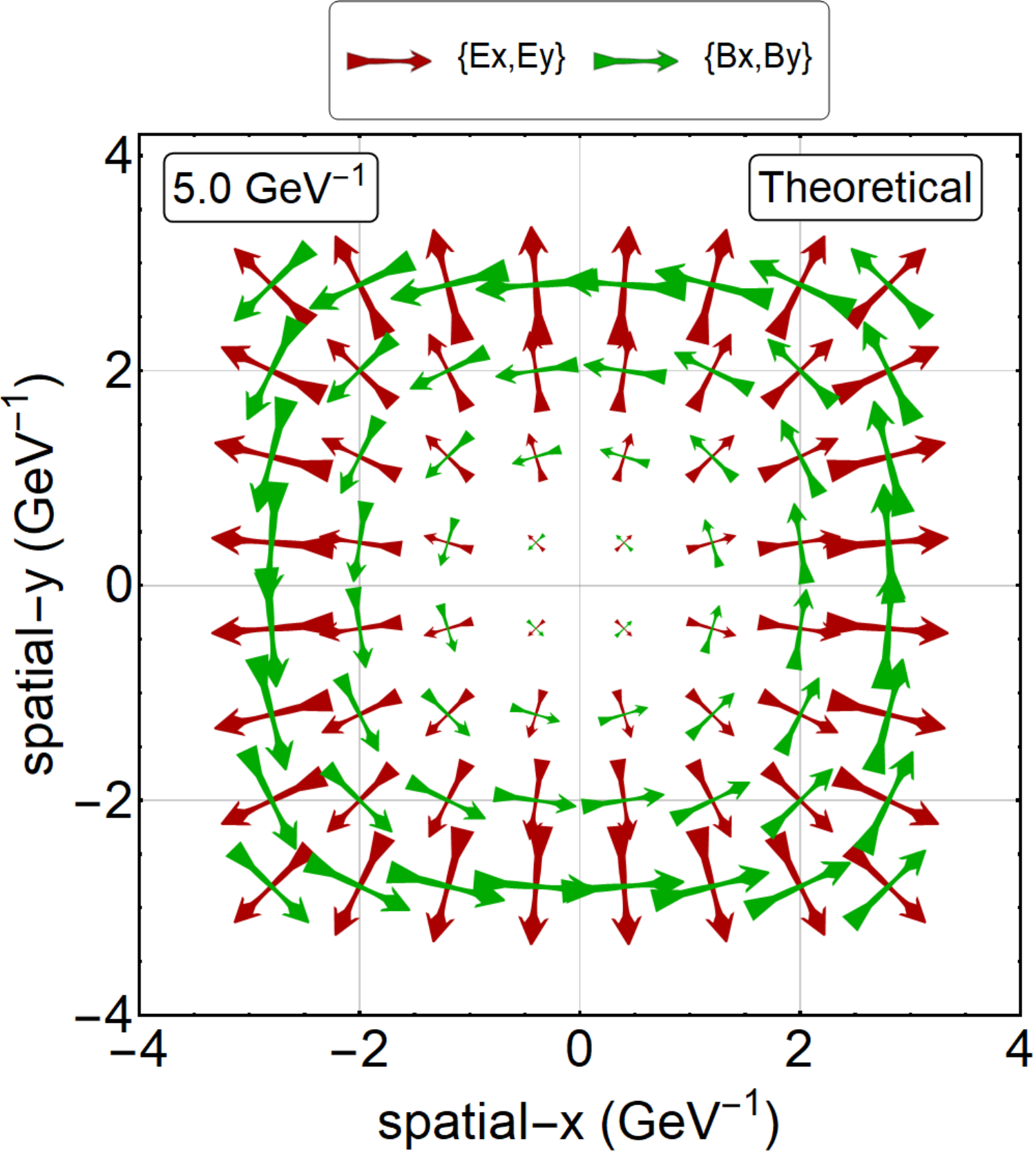} & \includegraphics[scale=0.25]{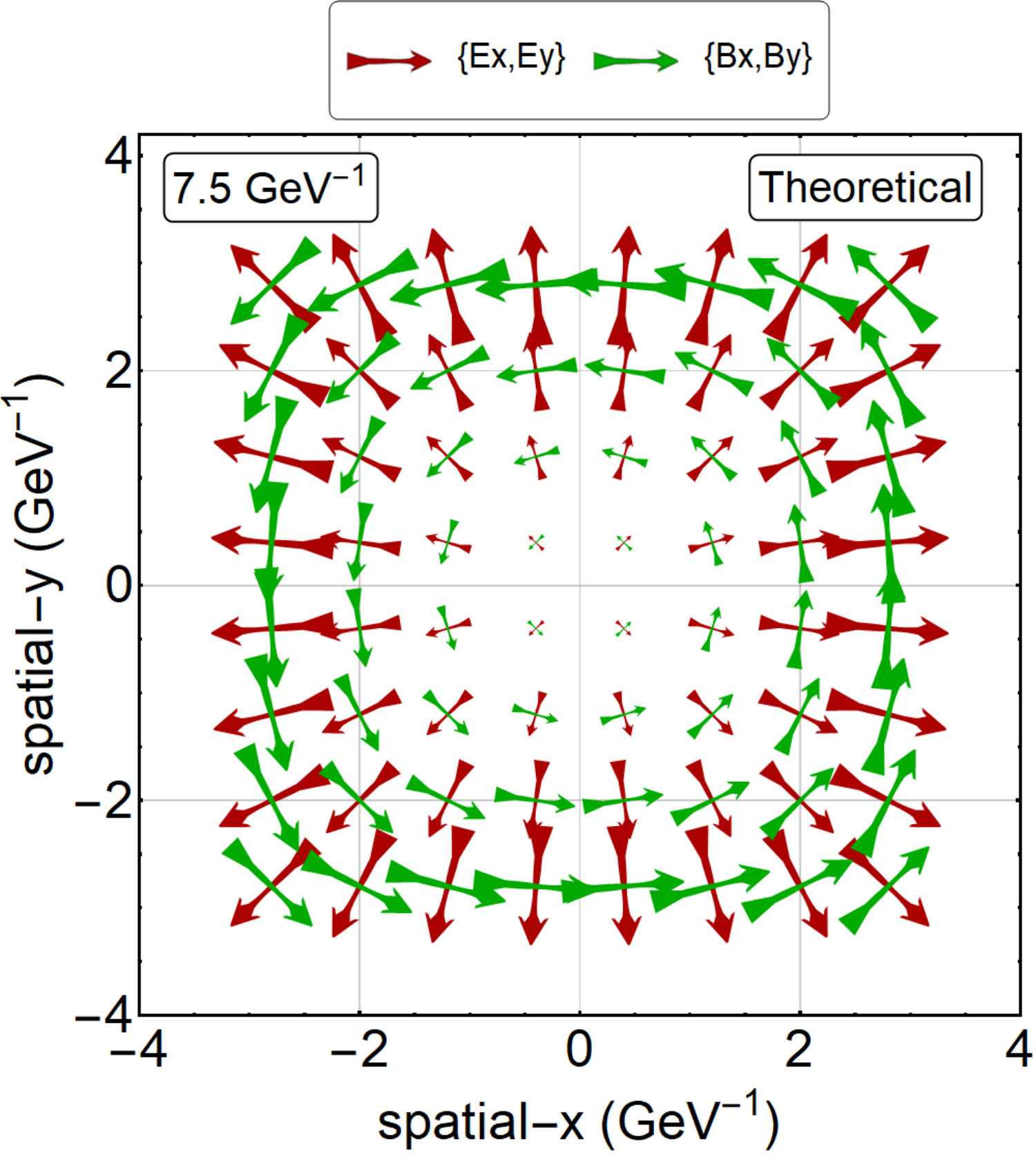}\tabularnewline
\includegraphics[scale=0.25]{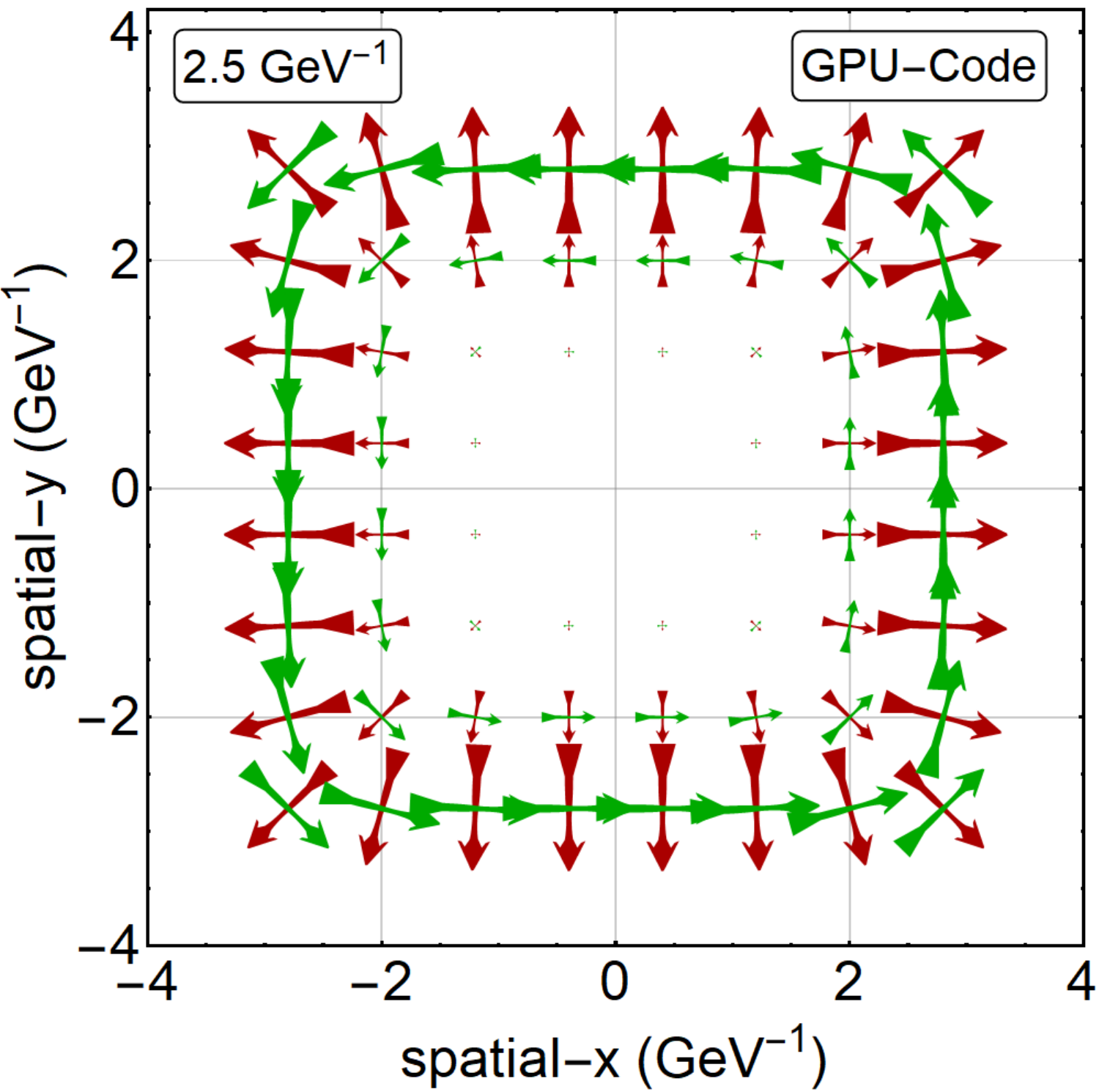} & \includegraphics[scale=0.25]{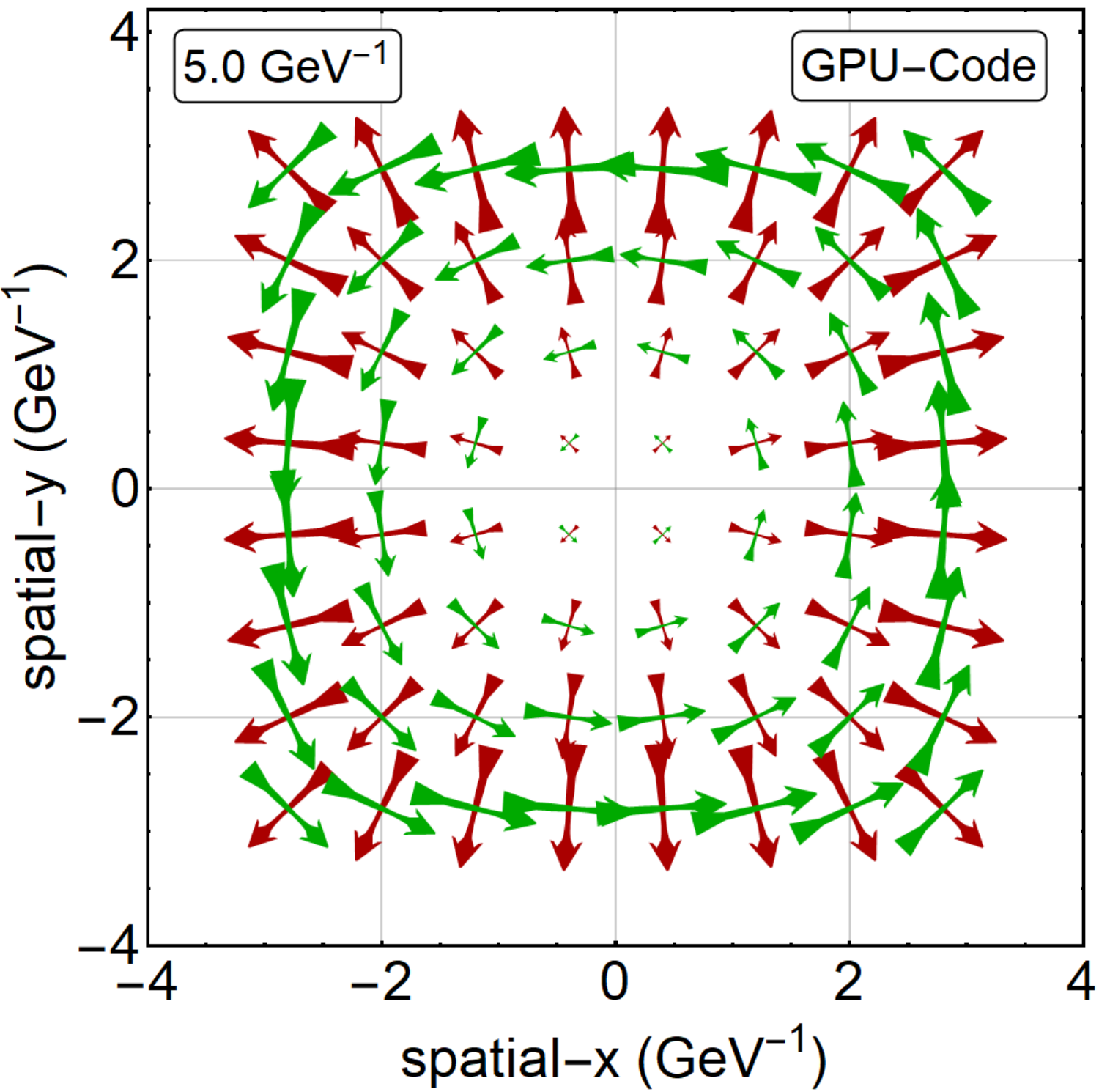} & \includegraphics[scale=0.25]{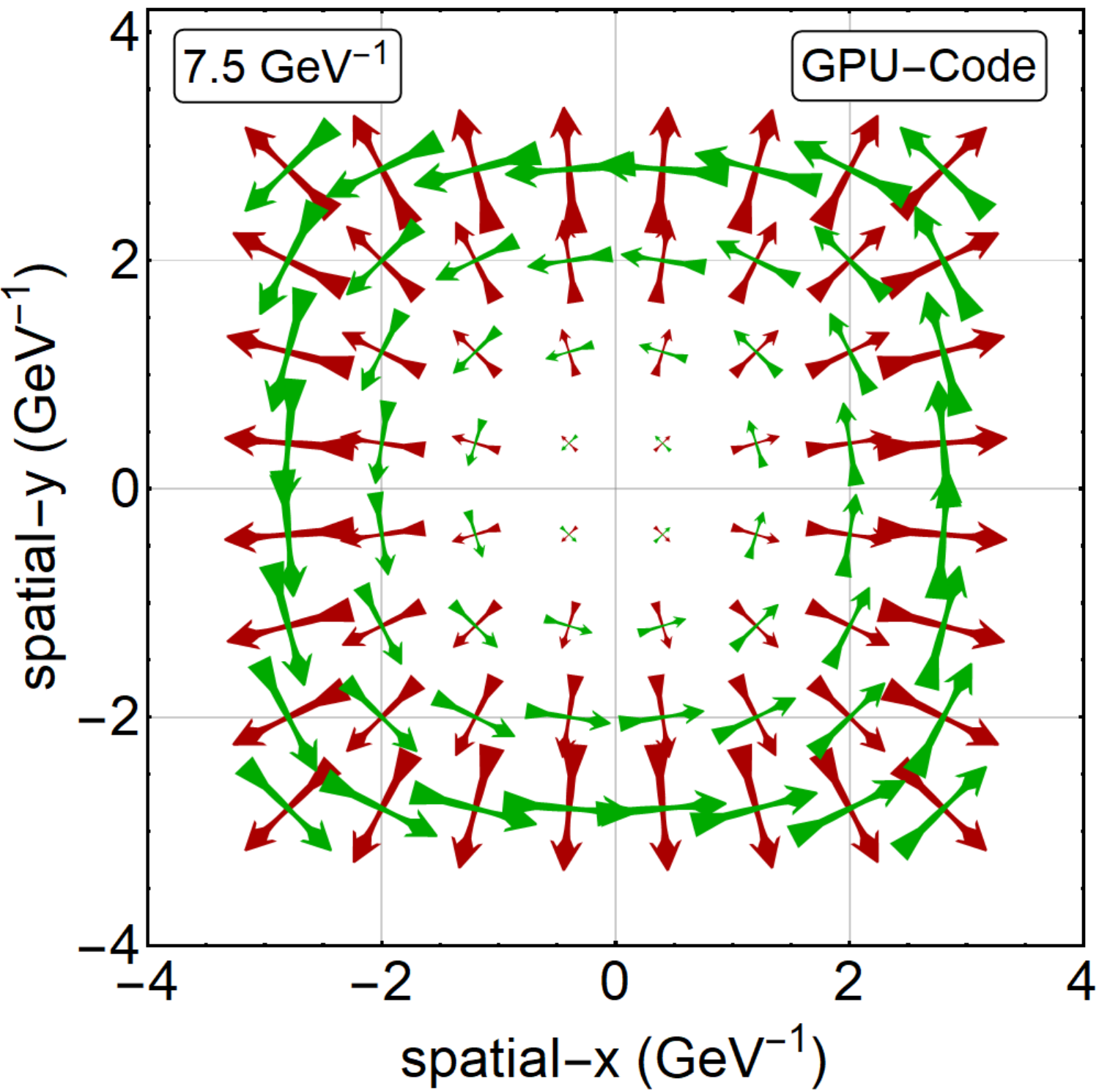}\tabularnewline
\end{tabular}
\par\end{centering}
\caption{Electric and magnetic field in the XOY plane with constant $\rho$
and $\mathbf{J}$. The upper panel gives the theoretical results via
Mathematica with built-in function NIntegrate, while the lower panel
gives the results obtained by the GPU code. \label{fig:Electric-and-magnetic}}
\end{figure}

\begin{figure}
\begin{centering}
\begin{tabular}{ccc}
\includegraphics[scale=0.25]{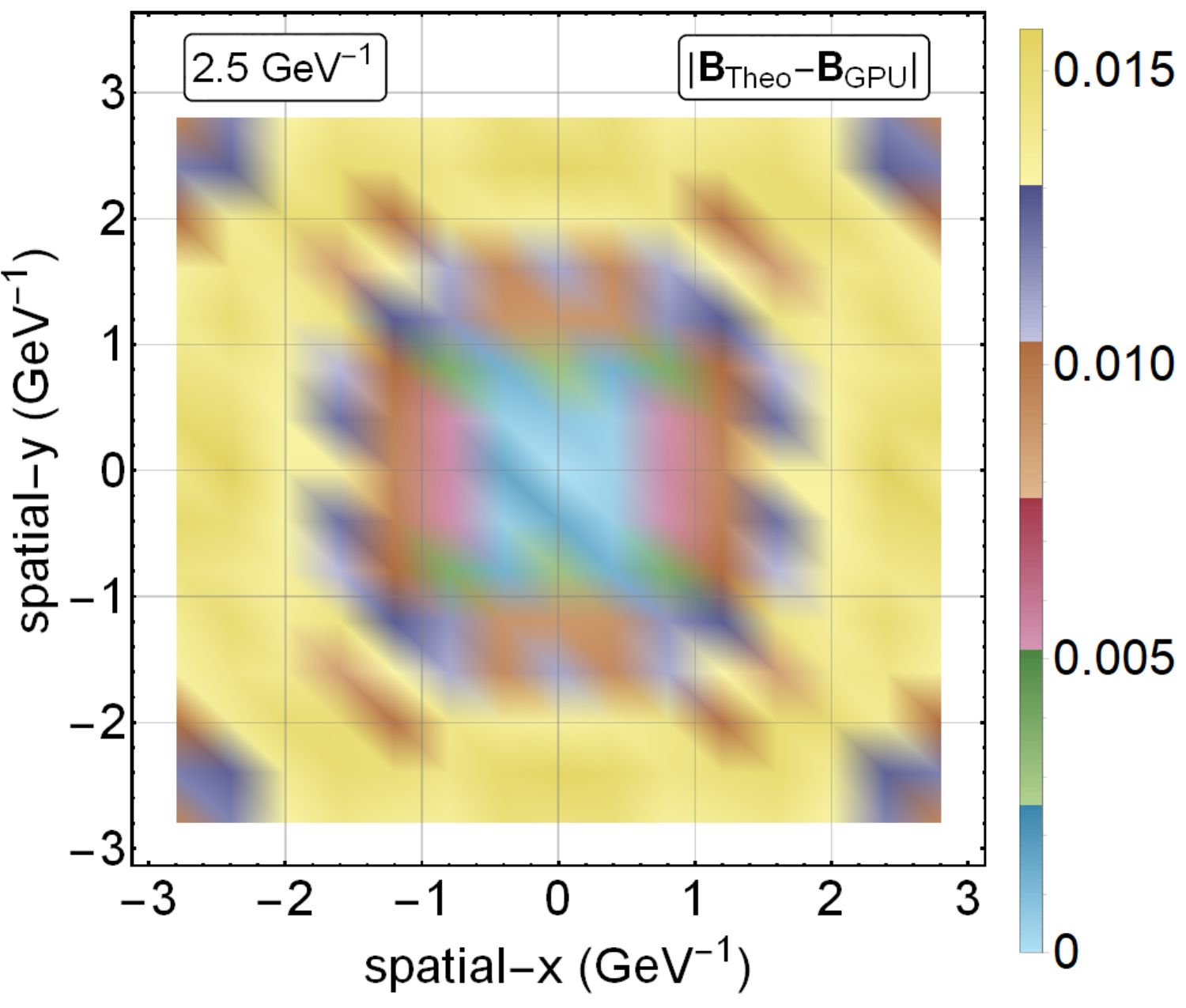} & \includegraphics[scale=0.25]{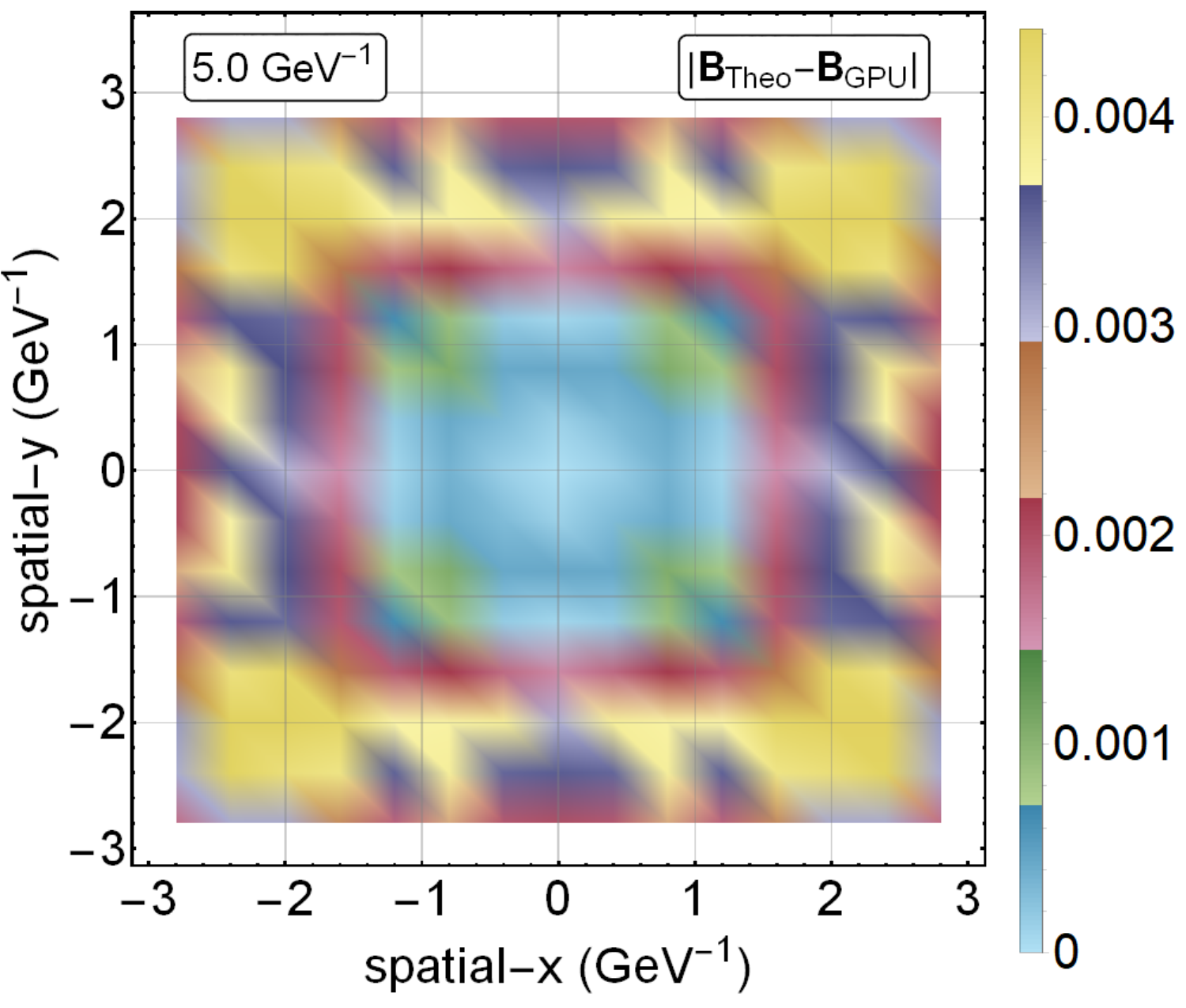} & \includegraphics[scale=0.25]{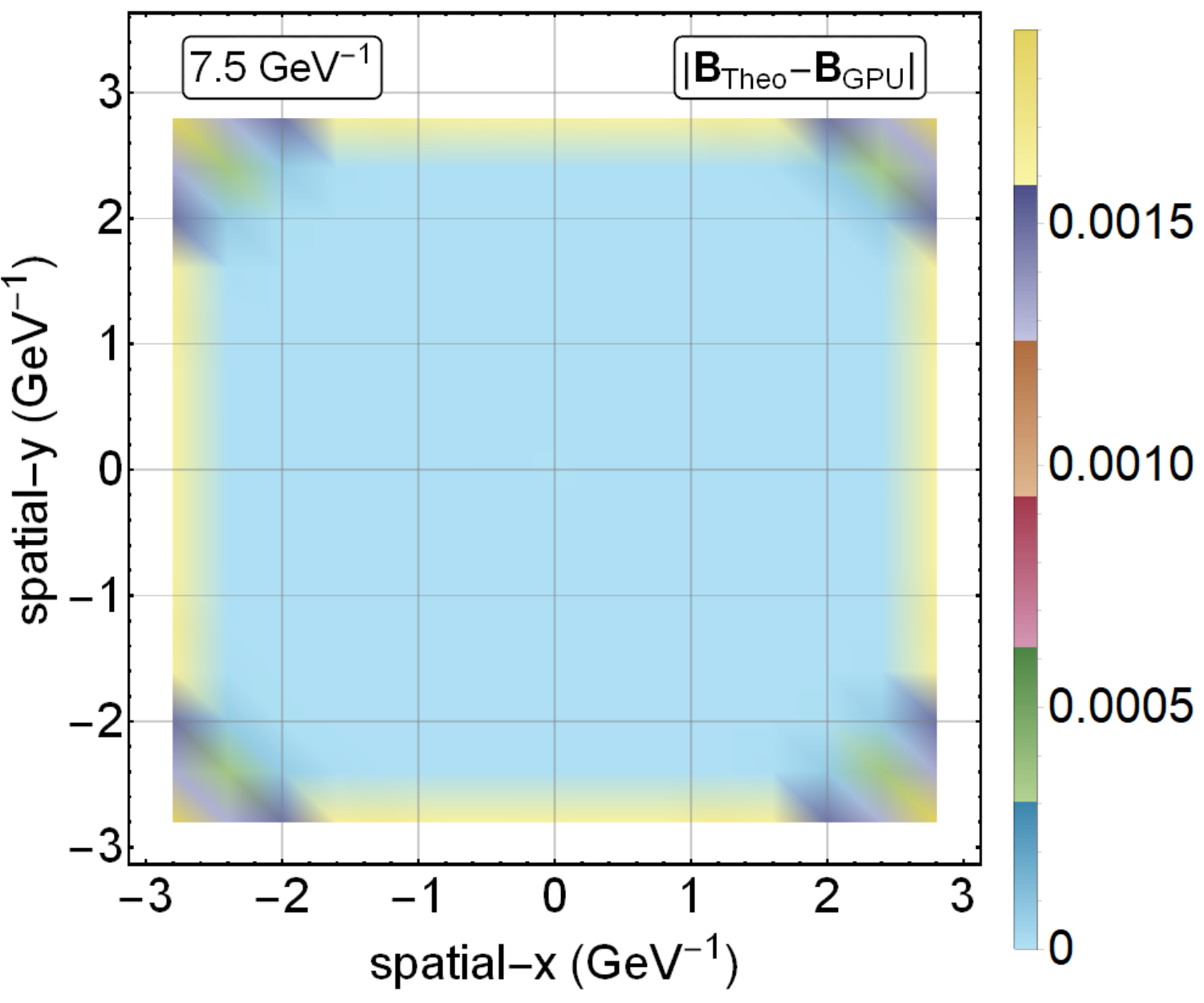}\tabularnewline
\includegraphics[scale=0.25]{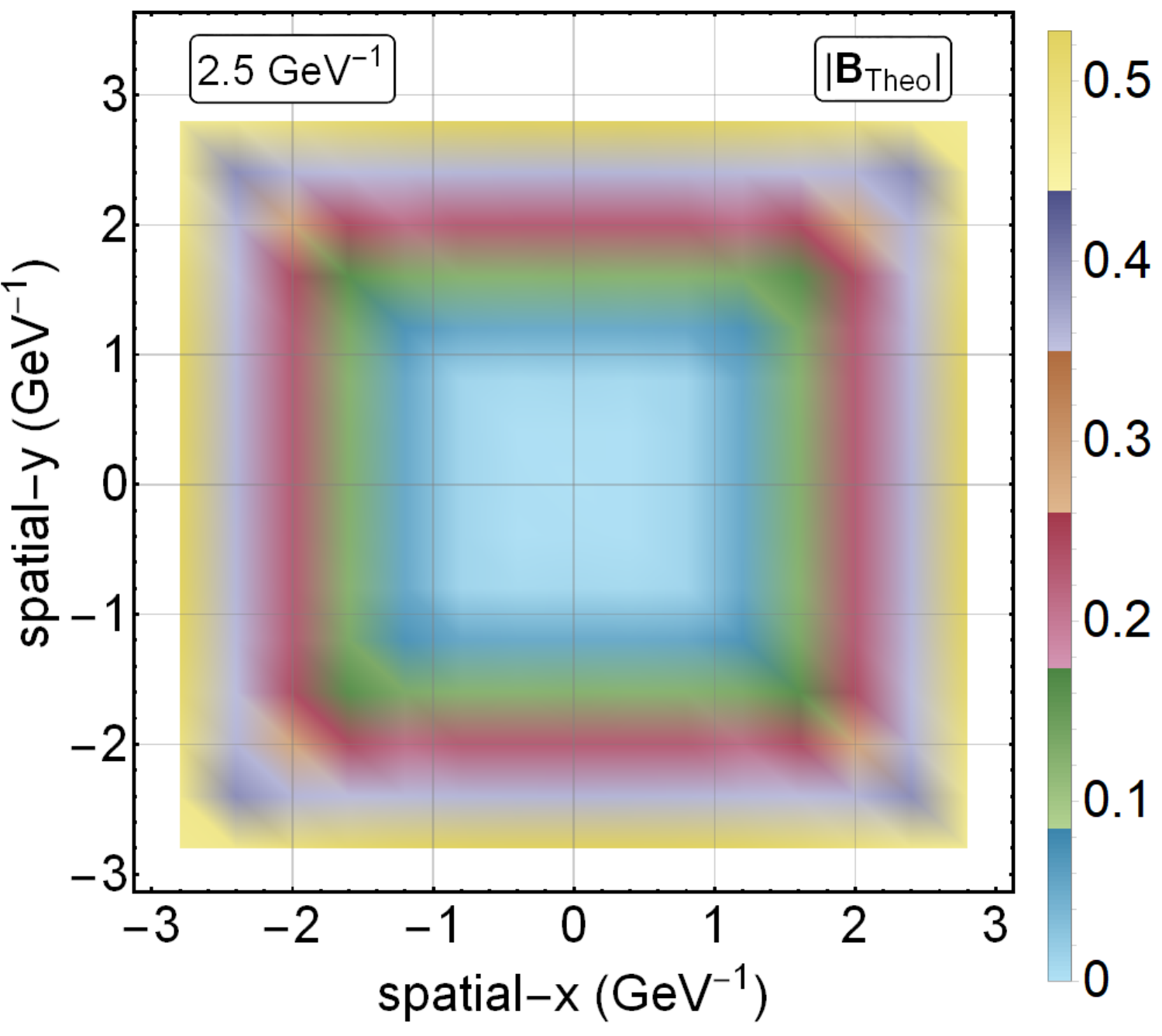} & \includegraphics[scale=0.25]{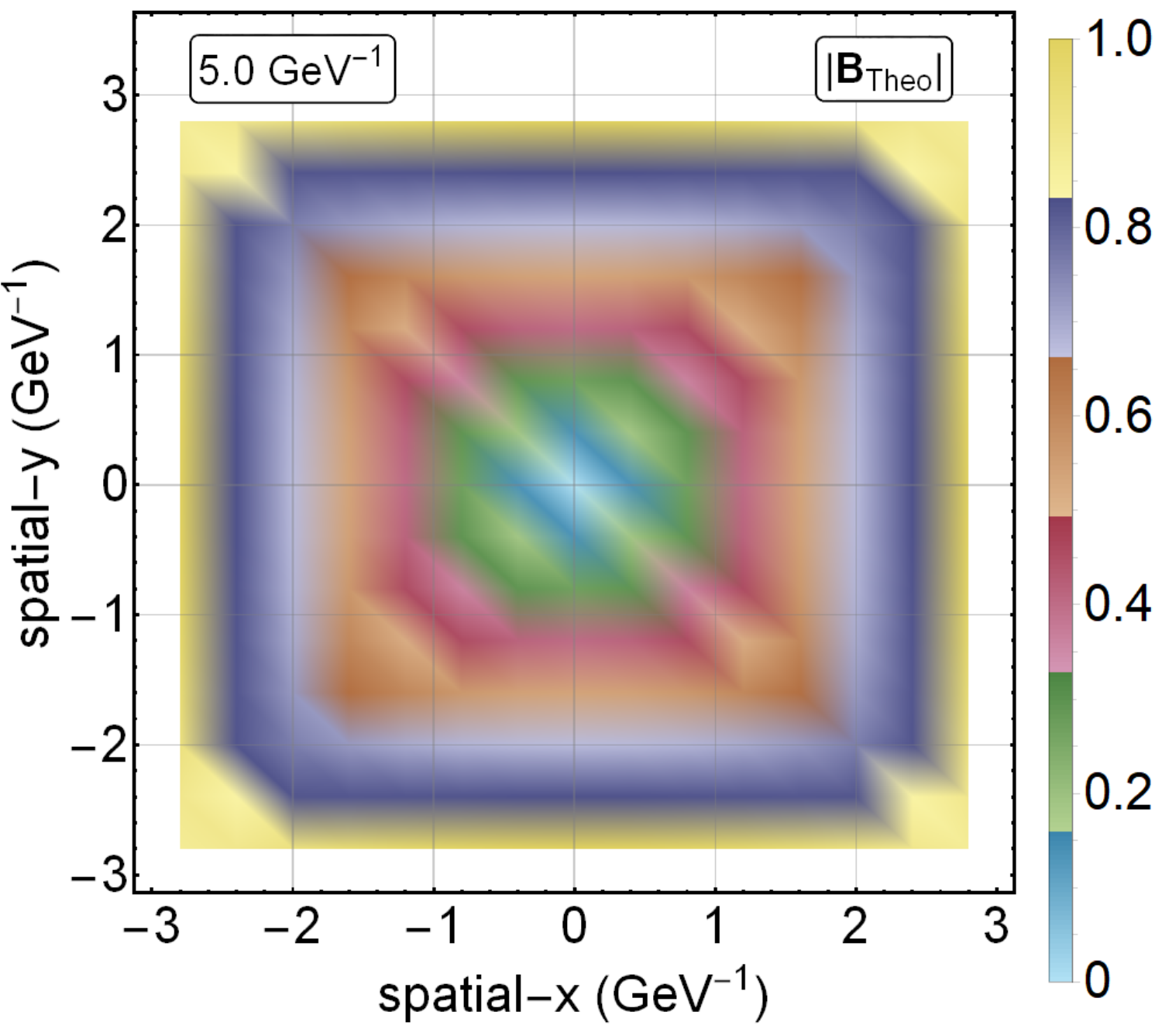} & \includegraphics[scale=0.25]{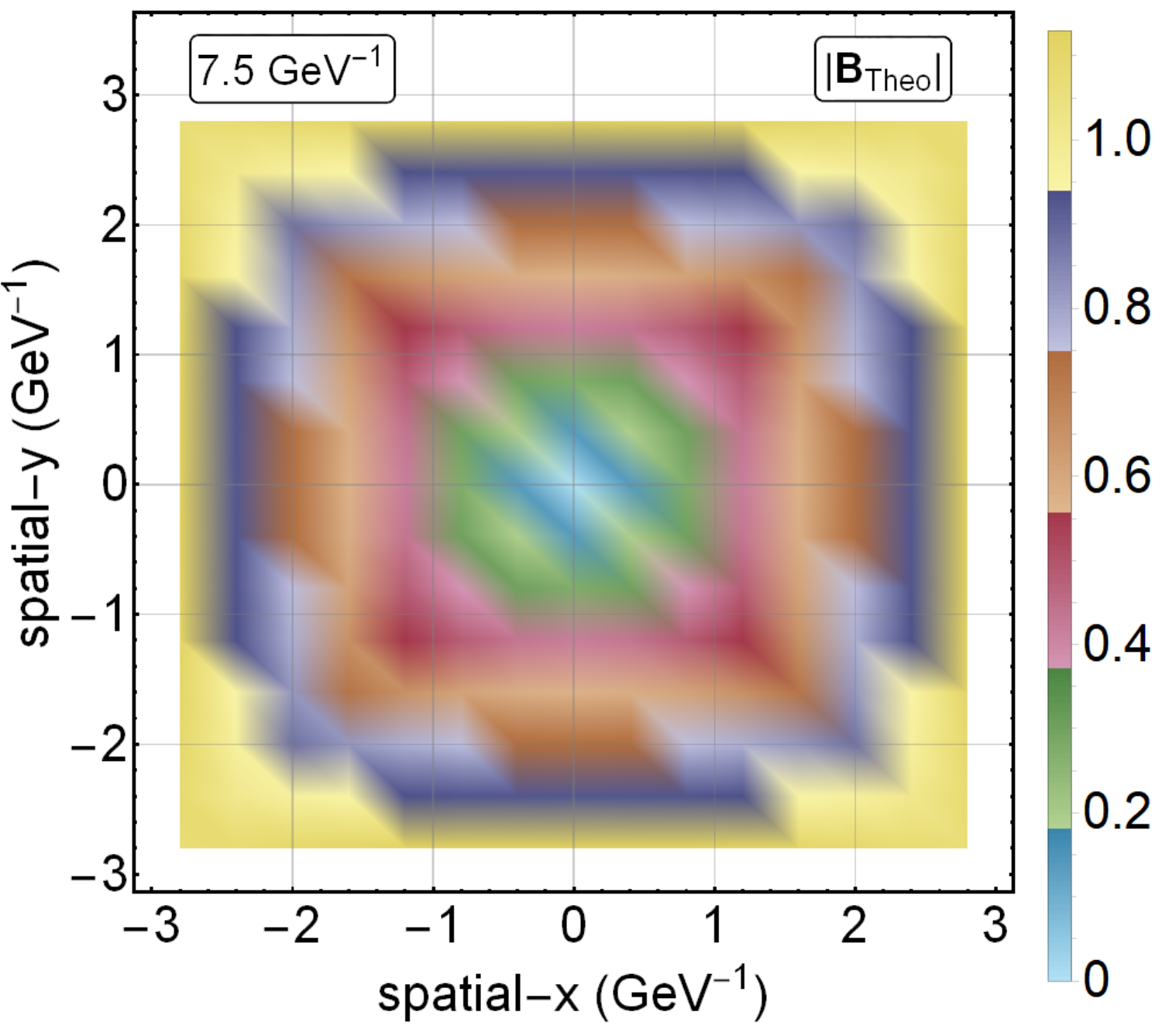}\tabularnewline
\end{tabular}
\par\end{centering}
\caption{Norm of $\mathbf{B}_{\text{Theo}}-\mathbf{B}_{\text{GPU}}$ and $\mathbf{B}_{\text{Theo}}$
with constant sources.\label{fig:Norm-of--1}}
\end{figure}

Figs. \ref{fig:Electric-and-magnetic}\textasciitilde\ref{fig:Norm-of--1}
show the results of the obtained electric and magnetic fields in the
XOY plane. The integration domain (also the domain of the observational
region) is of size $[-3\text{ GeV}^{-1},3\text{\ensuremath{\text{ GeV}^{-1}}}]^{3}$.
The theoretical results are obtained via NIntegrate in Mathematica
with screening length $\delta=0.2\text{ \ensuremath{\text{ GeV}^{-1}}}$,
which is half the grid length in the GPU-Code. In the GPU implementation
we have chosen $n_{x,\text{s}}=n_{y,\text{s}}=n_{z,\text{s}}=n_{x,\text{o}}=n_{y,\text{o}}=n_{z,\text{o}}=15$
such that each 3D grid is of volume size $(0.4\text{ \ensuremath{\text{ GeV}^{-1}}})^{3}$
.

Fig. \ref{fig:Electric-and-magnetic} compares $\mathbf{E}$ and $\mathbf{B}$
at time snapshots 2.5 $\text{GeV}^{-1}$, 5 $\text{GeV}^{-1}$ and
7.5 $\text{GeV}^{-1}$. The upper panel shows the theoretical results
with screening volume $\delta^{3}=(0.2\text{ \ensuremath{\text{GeV}^{-1}}})^{3}$,
and the lower panel gives the GPU results. We can see that the two
results follow the similar patterns at all time snapshots. Since the
sources are constant, we expect a saturation of $\mathbf{E}$ and
$\mathbf{B}$ at later times (e.g., when $t\geq5\text{\ensuremath{\text{ GeV}^{-1}}}$
the fields remain almost unchanged). The near saturation at around
5 $\text{GeV}^{-1}$ is reasonable since the longest transmission
length of the field is $6\text{ \ensuremath{\text{GeV}^{-1}}}\times\sqrt[3]{3}\sim8.7\text{\ensuremath{\text{ GeV}^{-1}}}$,.
Thus the system must reach a constant field distribution after $8.7\text{\ensuremath{\text{ GeV}^{-1}}}$.

\begin{figure}
\begin{centering}
\begin{tabular}{ccc}
\includegraphics[scale=0.25]{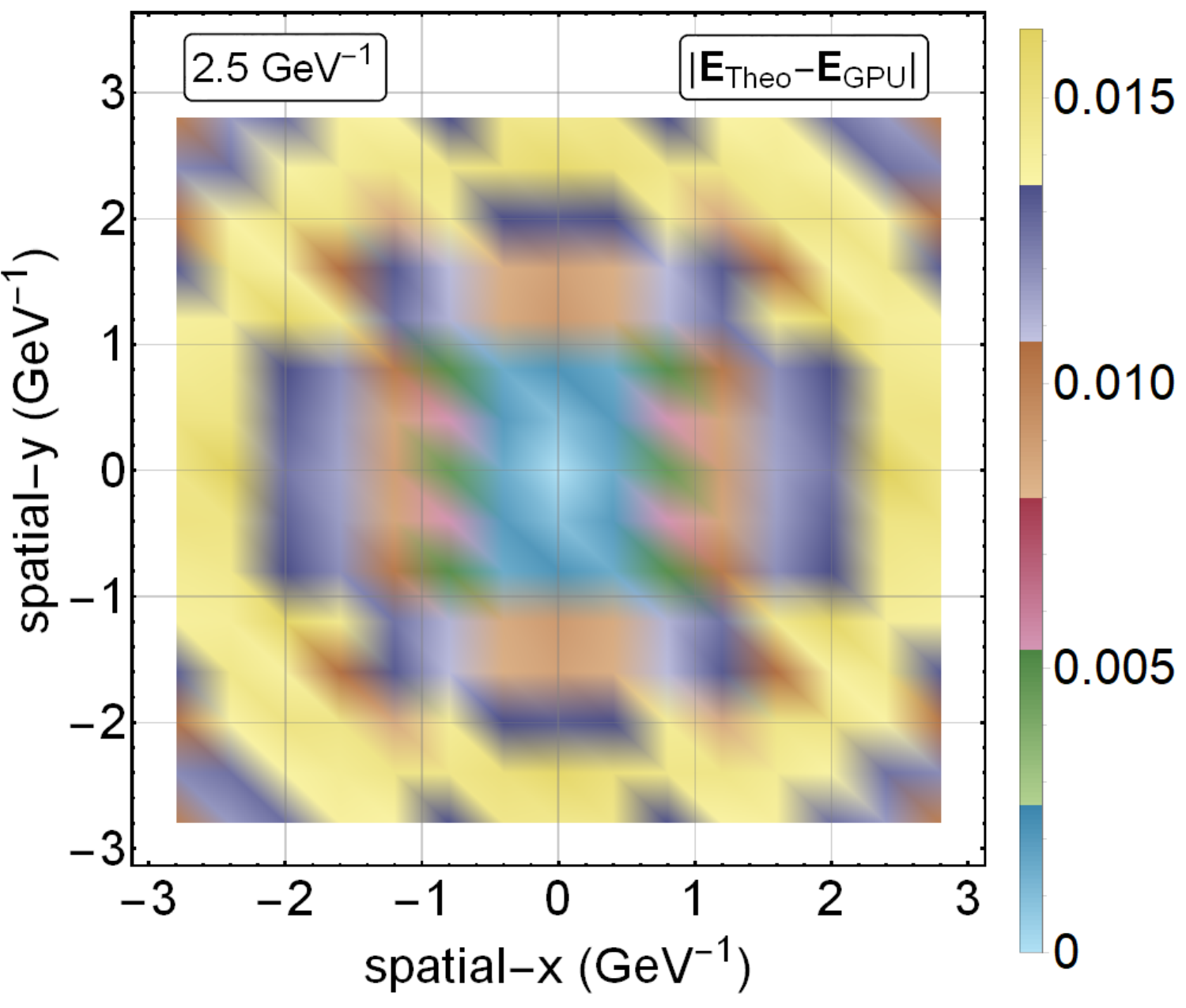} & \includegraphics[scale=0.25]{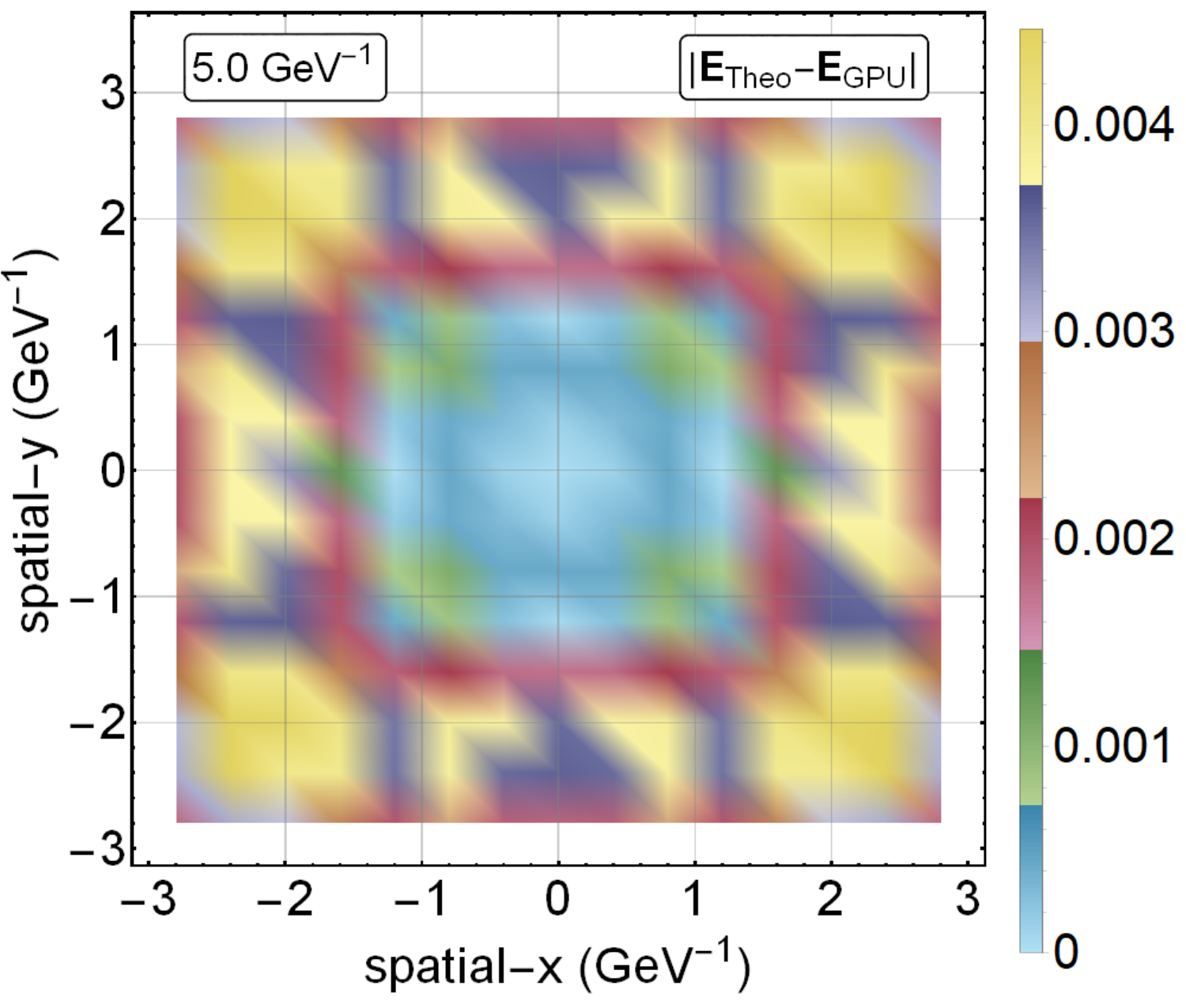} & \includegraphics[scale=0.25]{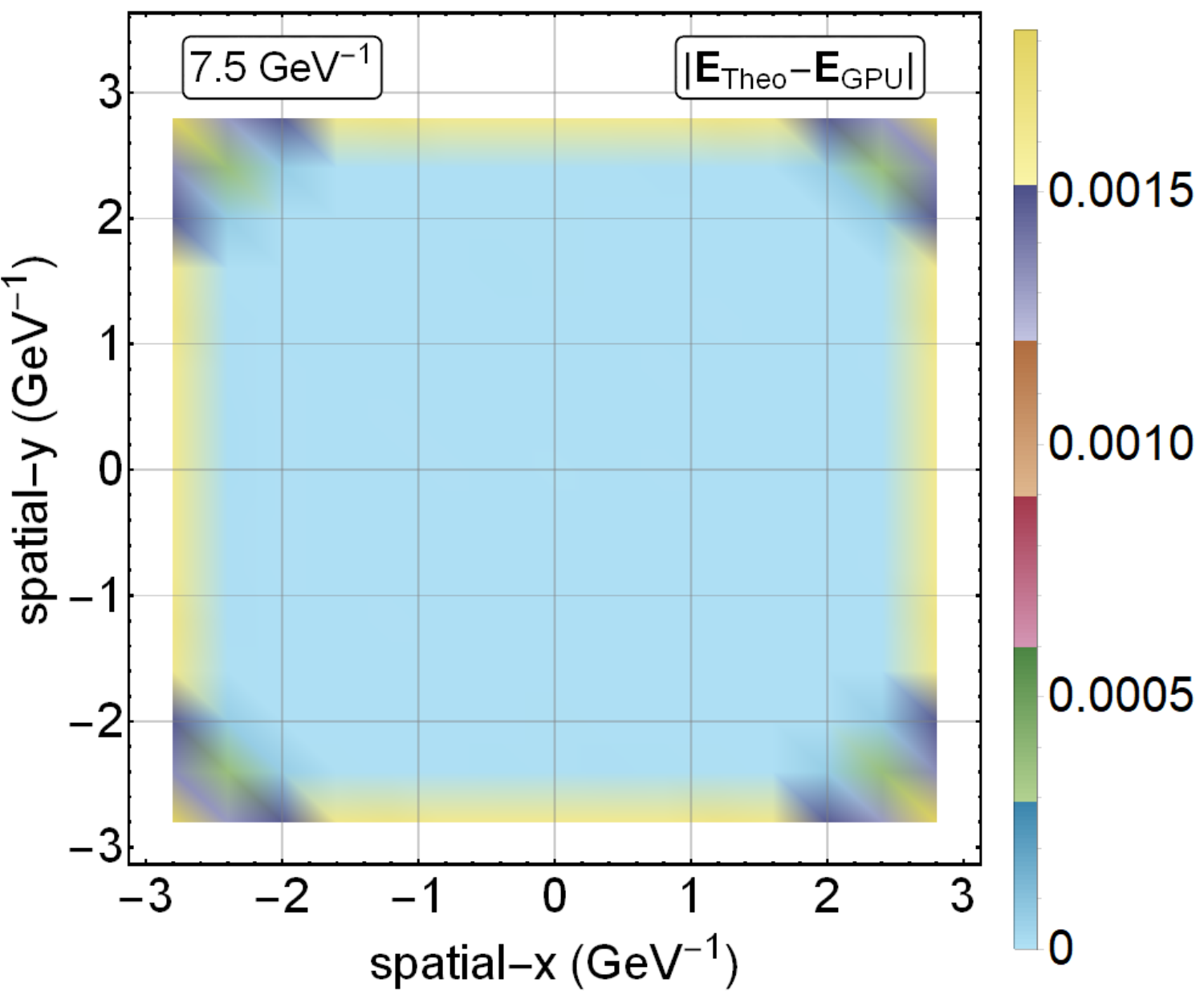}\tabularnewline
\includegraphics[scale=0.25]{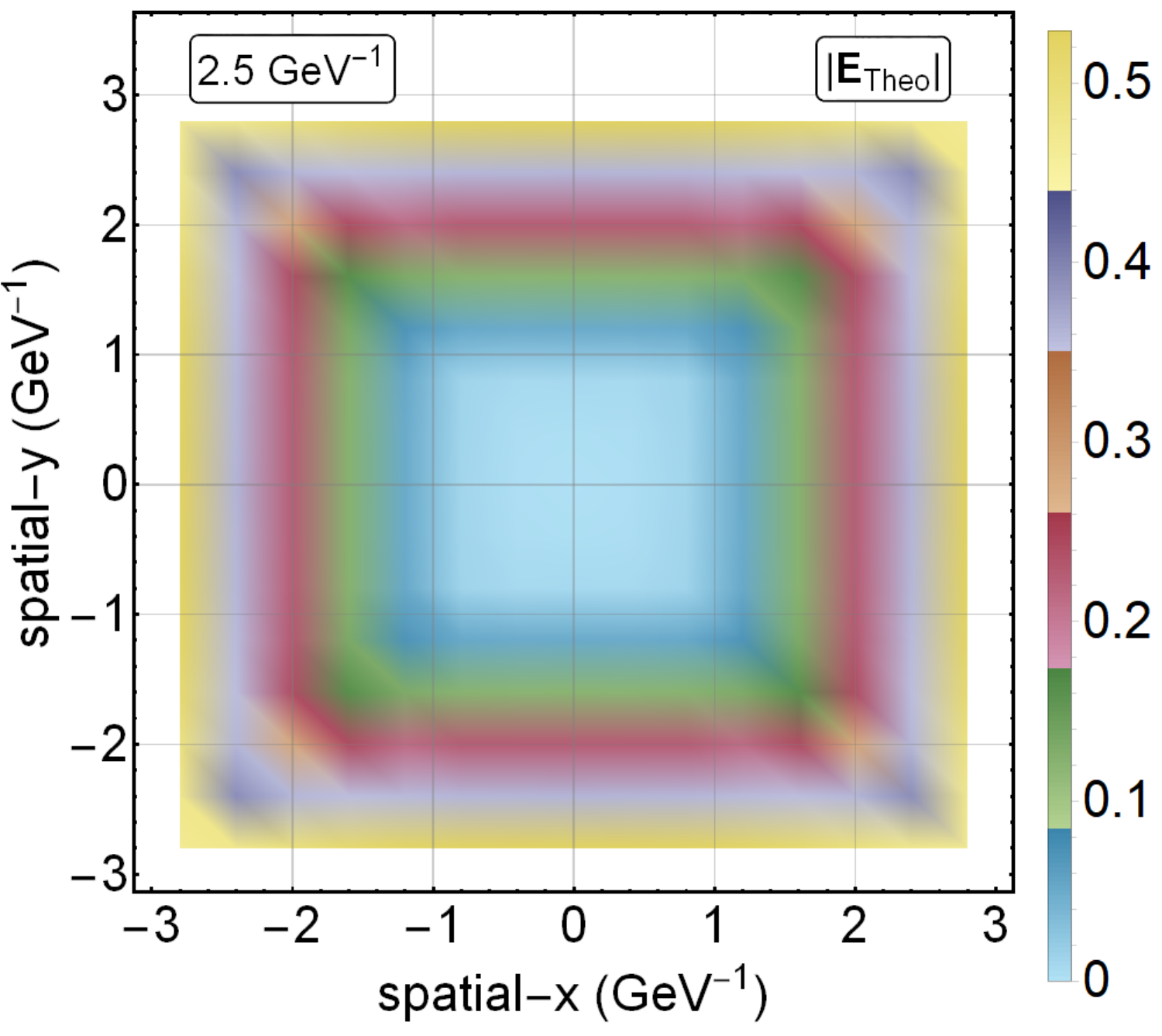} & \includegraphics[scale=0.25]{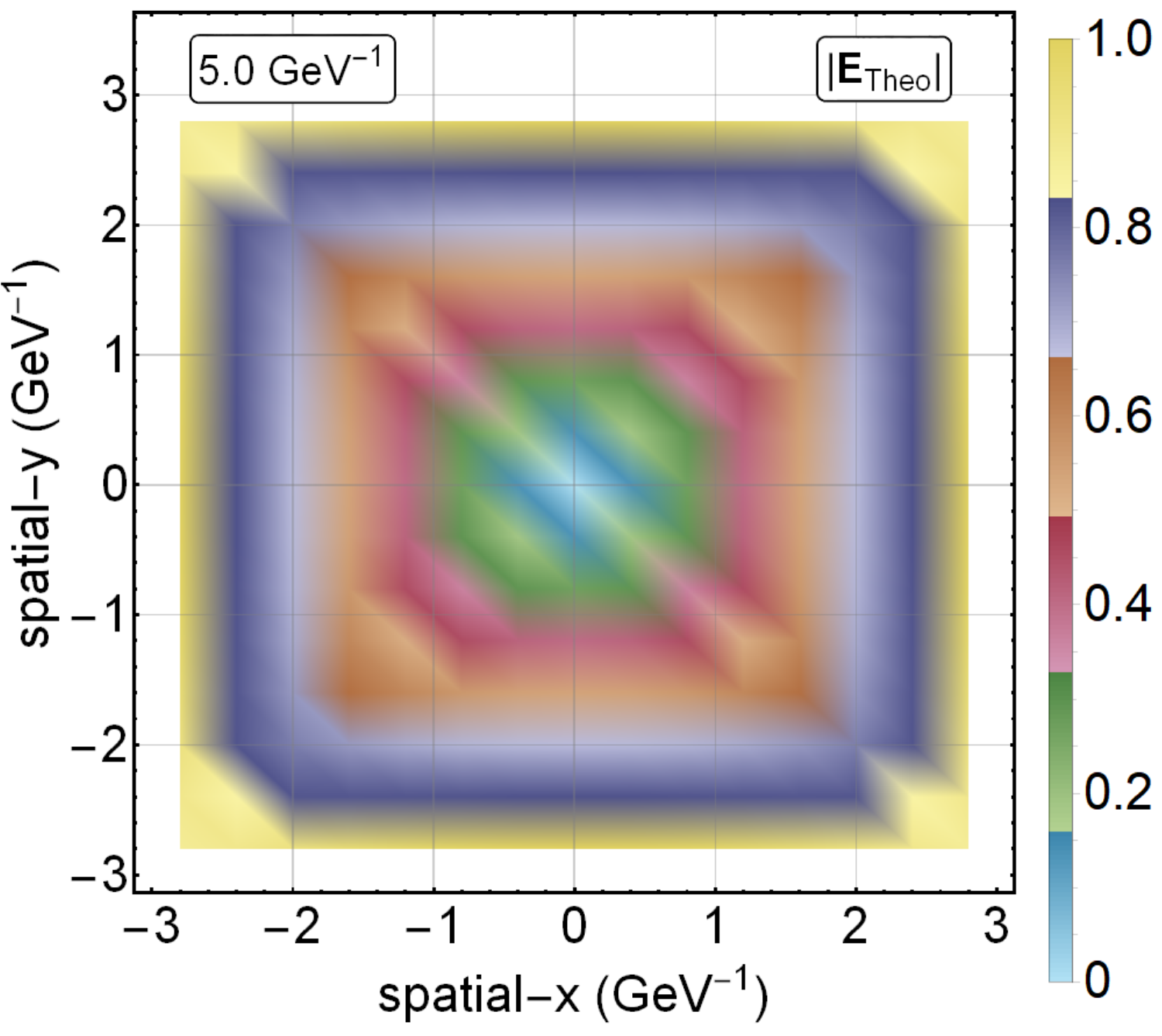} & \includegraphics[scale=0.25]{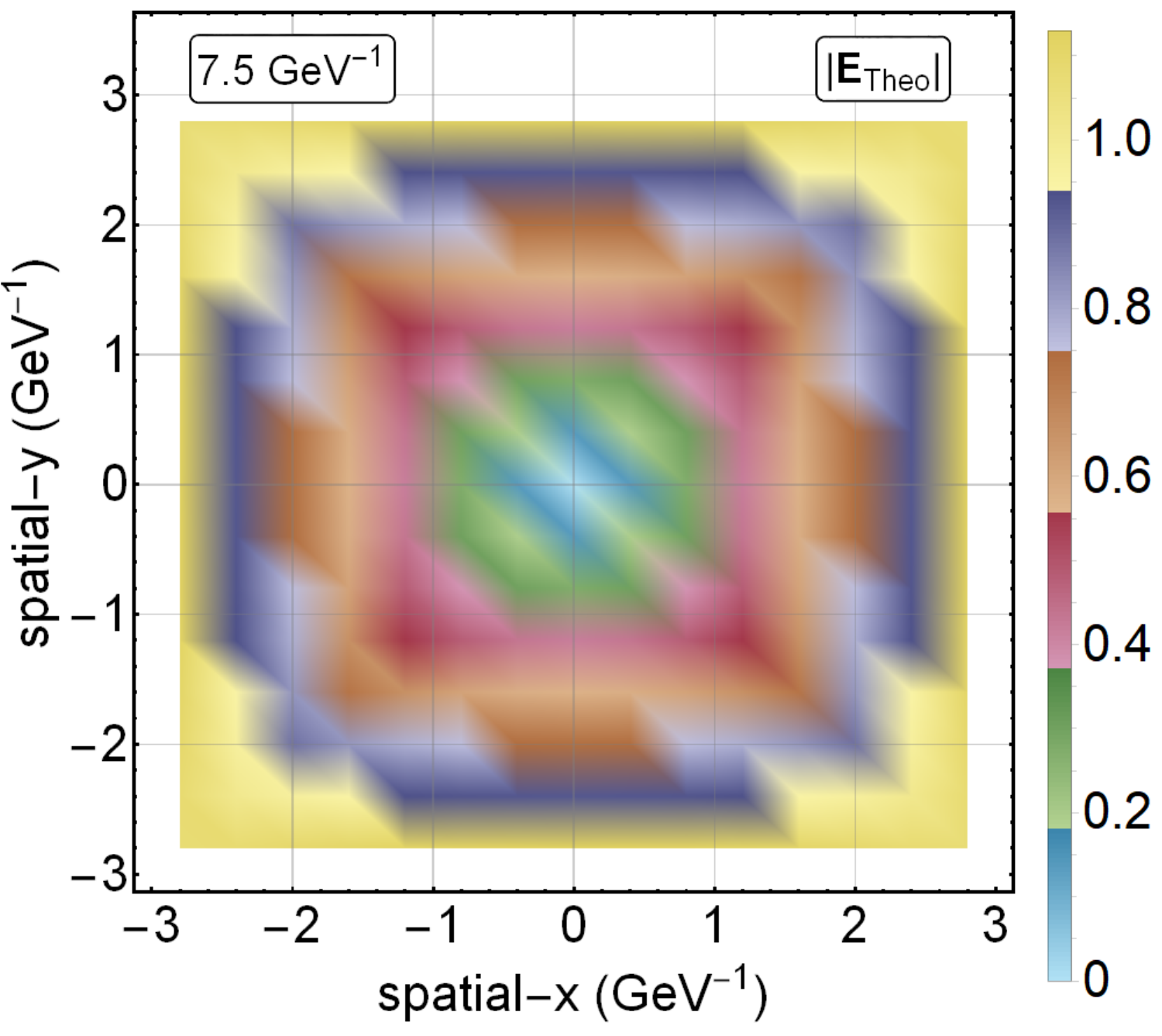}\tabularnewline
\end{tabular}
\par\end{centering}
\caption{Norm of $\mathbf{E}_{\text{Theo}}-\mathbf{E}_{\text{GPU}}$ and $\mathbf{E}_{\text{Theo}}$
with constant sources. The norm of the two vectors are defined in
the XOY plane, i.e., $|\mathbf{A}|\equiv\sqrt{A_{x}^{2}+A_{y}^{2}}$.\label{fig:Norm-of-}}
\end{figure}

Fig. \ref{fig:Norm-of-} and \ref{fig:Norm-of--1} give the comparison
of $|\mathbf{E}_{\text{Theo}}-\mathbf{E}_{\text{GPU}}|$ with $|\mathbf{E}_{\text{Theo}}|$
and $|\mathbf{B}_{\text{Theo}}-\mathbf{B}_{\text{GPU}}|$ with $|\mathbf{B}_{\text{Theo}}|$.
In the XOY plane, the norms are taken as
\begin{eqnarray}
|\mathbf{E}_{\text{Theo}}-\mathbf{E}_{\text{GPU}}| & = & \left[(E_{x,\text{Theo}}-E_{x,\text{GPU}})^{2}\right.\nonumber \\
 &  & +\left.(E_{y,\text{Theo}}-E_{y,\text{GPU}})^{2}\right]^{1/2}\nonumber \\
|\mathbf{E}_{\text{Theo}}| & = & \sqrt{E_{x,\text{Theo}}^{2}+E_{y,\text{Theo}}^{2}}\nonumber \\
|\mathbf{B}_{\text{Theo}}-\mathbf{B}_{\text{GPU}}| & = & \left[(B_{x,\text{Theo}}-B_{x,\text{GPU}})^{2}\right.\nonumber \\
 &  & +\left.(B_{y,\text{Theo}}-B_{y,\text{GPU}})^{2}\right]^{1/2}\nonumber \\
|\mathbf{B}_{\text{Theo}}| & = & \sqrt{B_{x,\text{Theo}}^{2}+B_{y,\text{Theo}}^{2}},\label{eq:norms}
\end{eqnarray}
where we neglect the $z-$component of the fields for numerical convenience.
From the legend bars in Fig. \ref{fig:Norm-of-} and \ref{fig:Norm-of--1},
we can see that the maximum deviation of the GPU-Code from the theoretical
results is within $0.015/0.5\sim3%\ensuremath{\%}
$ for both $\mathbf{E}$ and $\mathbf{B}$.

\subsection{Sinusoidal sources\label{subsec:Sinusoidal-sources}}

In the sinusoidal case, we choose two different domains for the observational
($[-3\text{ GeV}^{-1},3\text{\ensuremath{\text{ GeV}^{-1}}}]^{2}\times[10\text{ GeV}^{-1},16\text{ GeV}^{-1}]$,
where $[10\text{ GeV}^{-1},16\text{ GeV}^{-1}]$ is the interval in
$z$ direction) and source regions ($[-3\text{ GeV}^{-1},3\text{\ensuremath{\text{ GeV}^{-1}}}]^{3}$).
The sources of sinusoidal $\rho$ and $\mathbf{J}$ take the following
form
\begin{eqnarray}
\rho(x,y,z,t) & = & \begin{cases}
0 & \text{if }t<0\\
3(\text{cos}(t)-1)\text{cos}(x+y+z) & \text{if }t\geq0
\end{cases}\nonumber \\
\mathbf{J}(x,y,z,t) & = & \begin{cases}
\mathbf{0} & \text{if }t<0\\
(j_{x},j_{y},j_{z}) & \text{if }t\geq0
\end{cases},\label{eq:sin_sources}
\end{eqnarray}
where $j_{x}=j_{y}=j_{z}=\text{sin}(x+y+z)\text{sin}(t)$. Eq. (\ref{eq:sin_sources})
satisfies the continuum relation $\nabla\cdot\mathbf{J}+\partial\rho/\partial t=0$.

\begin{figure}
\begin{centering}
\begin{tabular}{ccc}
\includegraphics[scale=0.25]{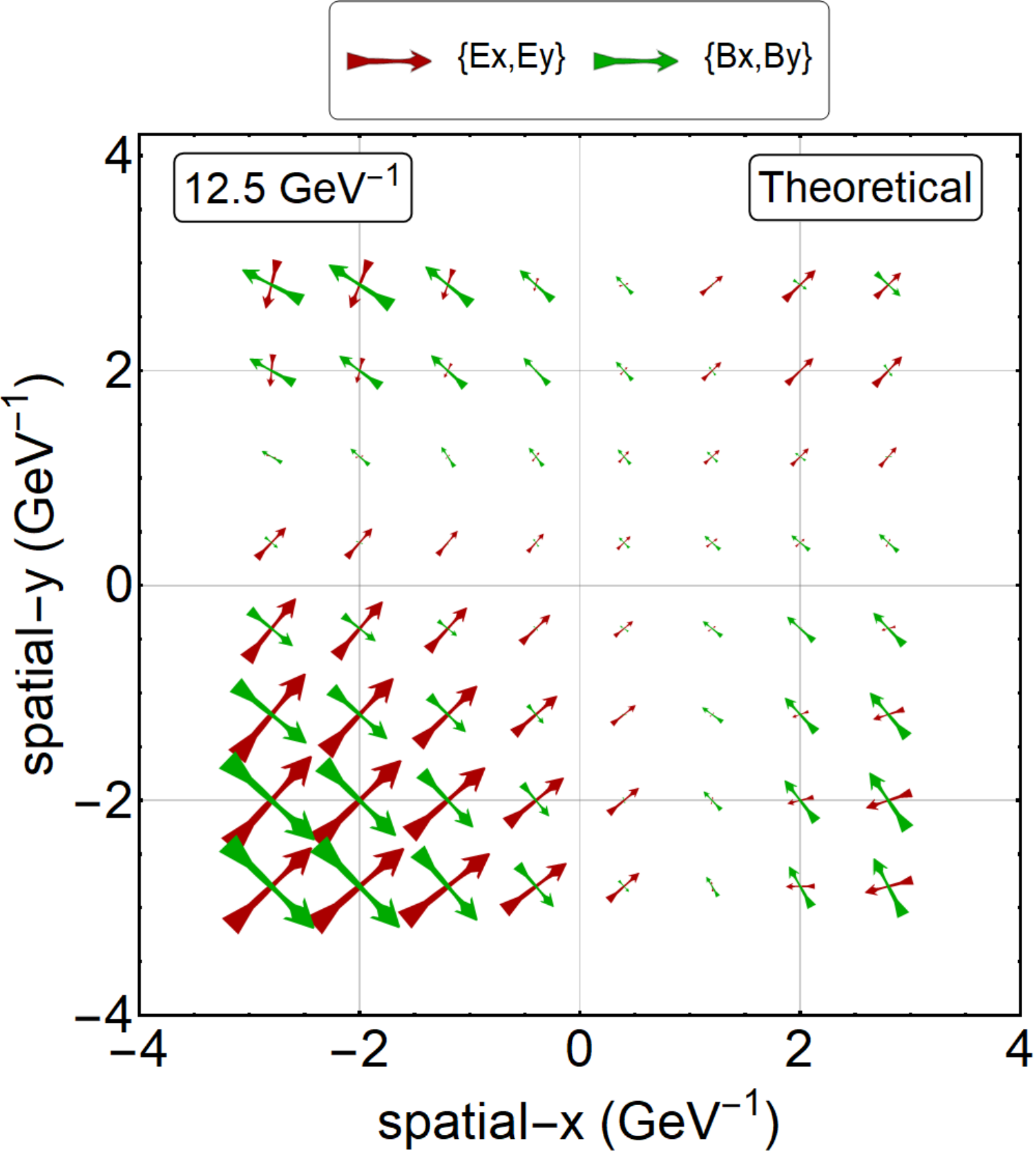} & \includegraphics[scale=0.25]{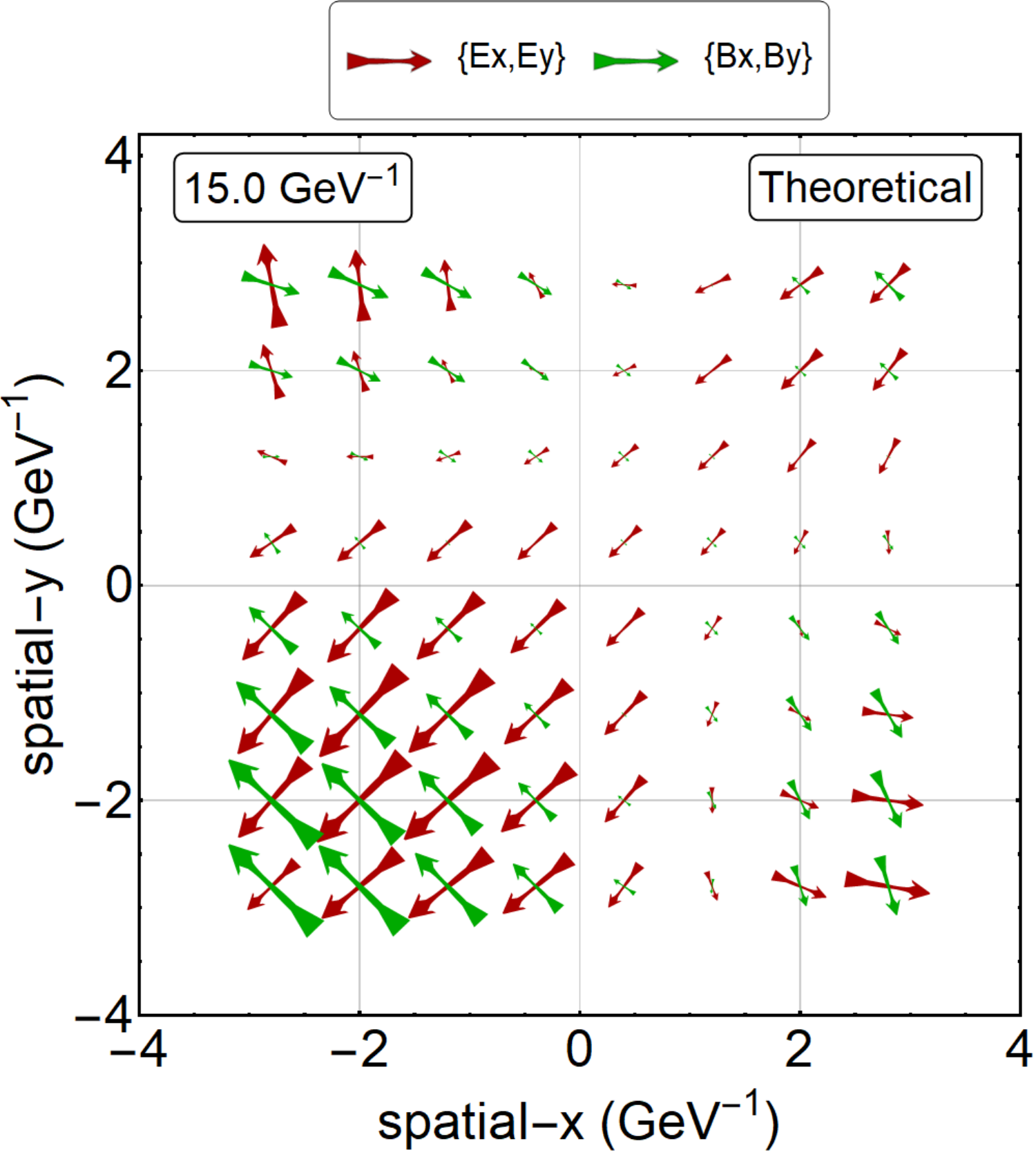} & \includegraphics[scale=0.25]{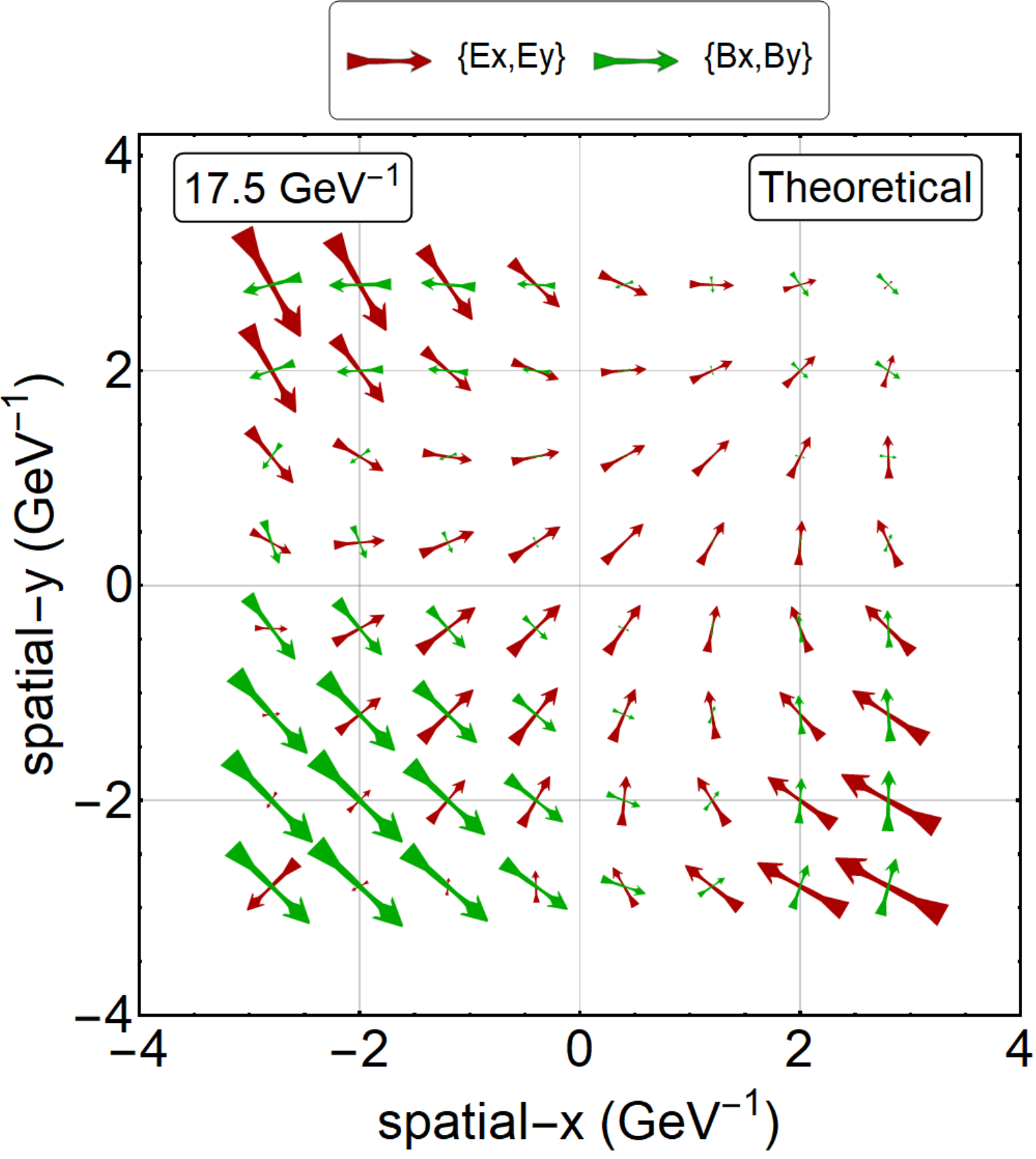}\tabularnewline
\includegraphics[scale=0.25]{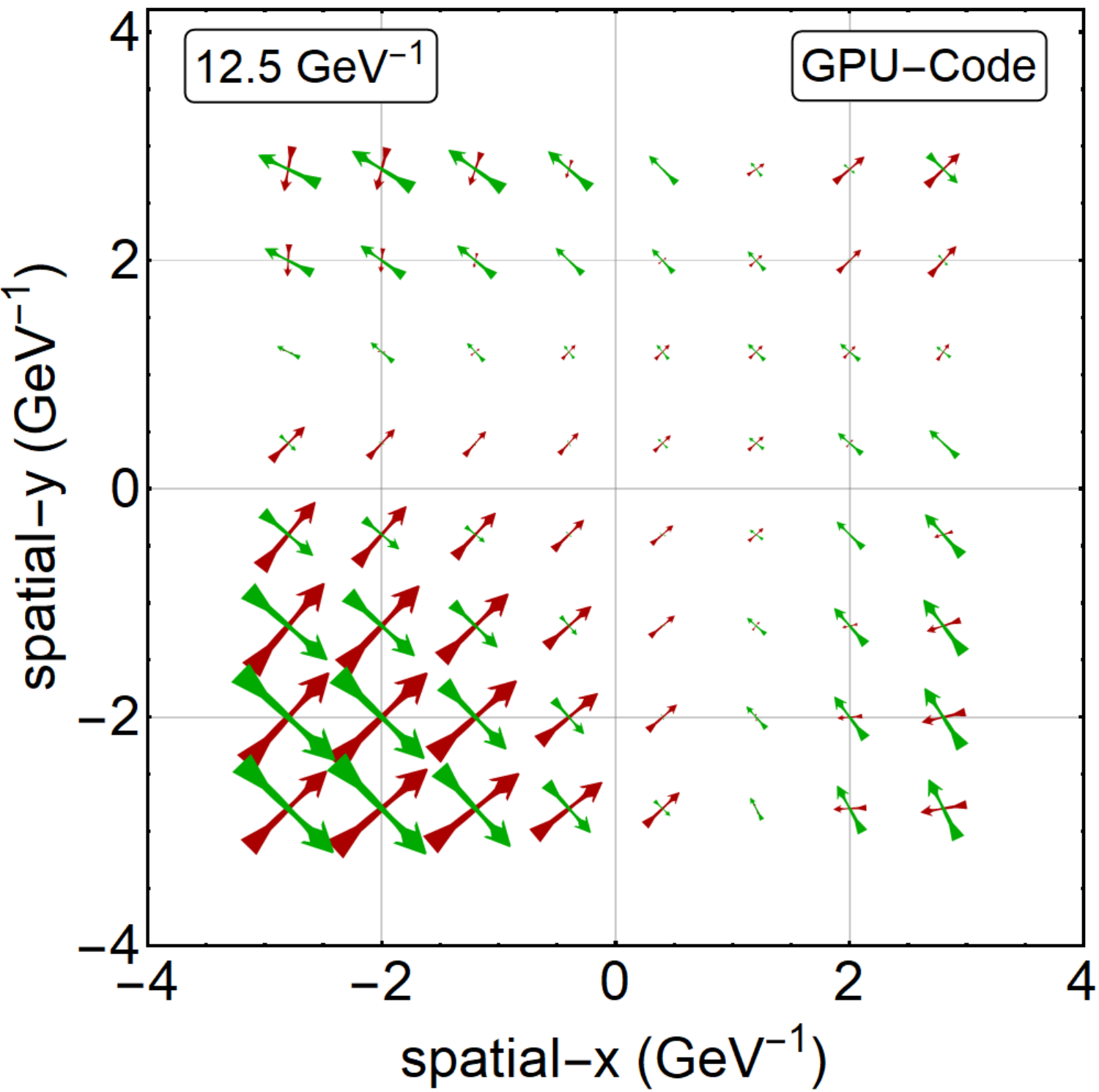} & \includegraphics[scale=0.25]{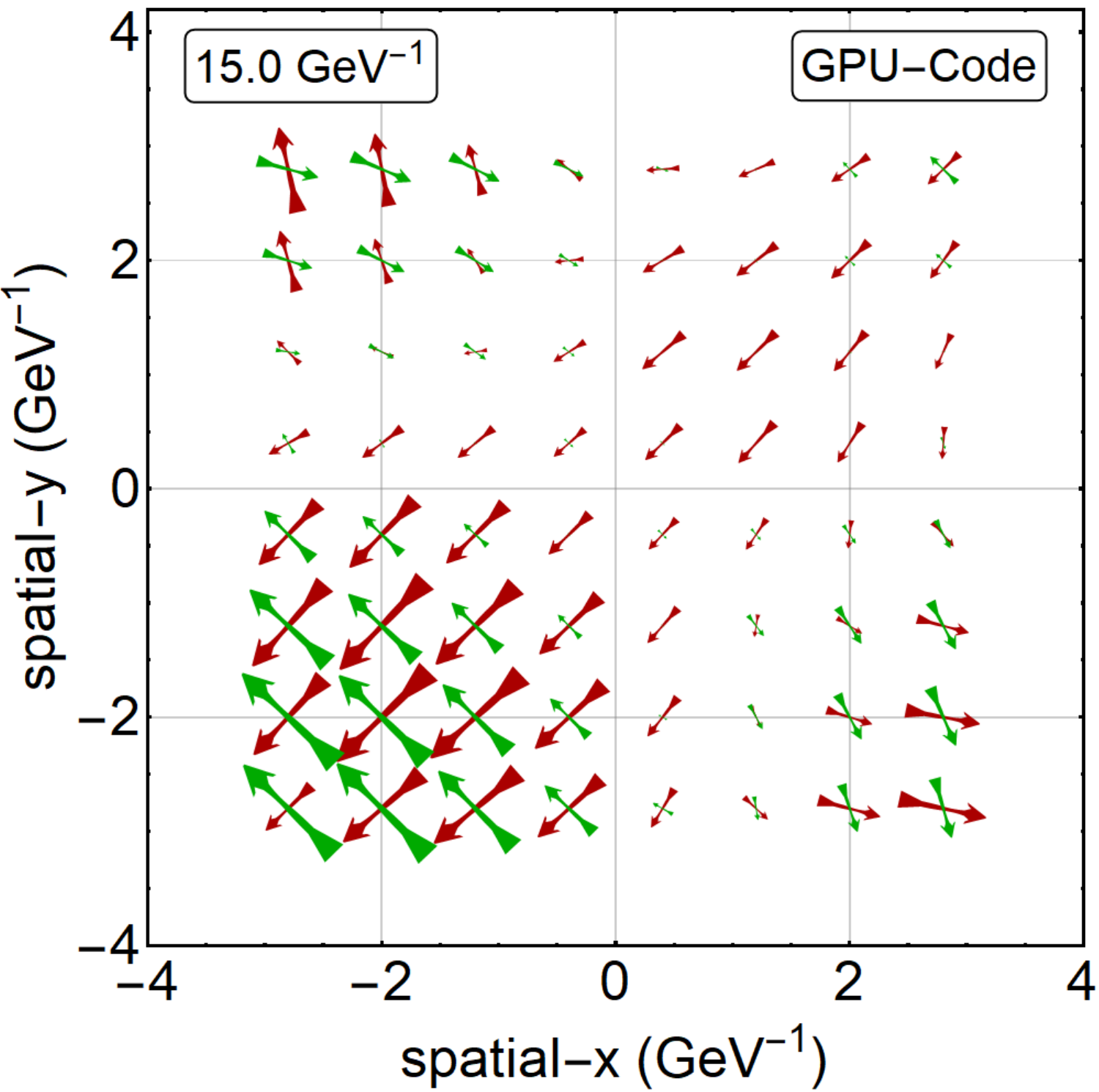} & \includegraphics[scale=0.25]{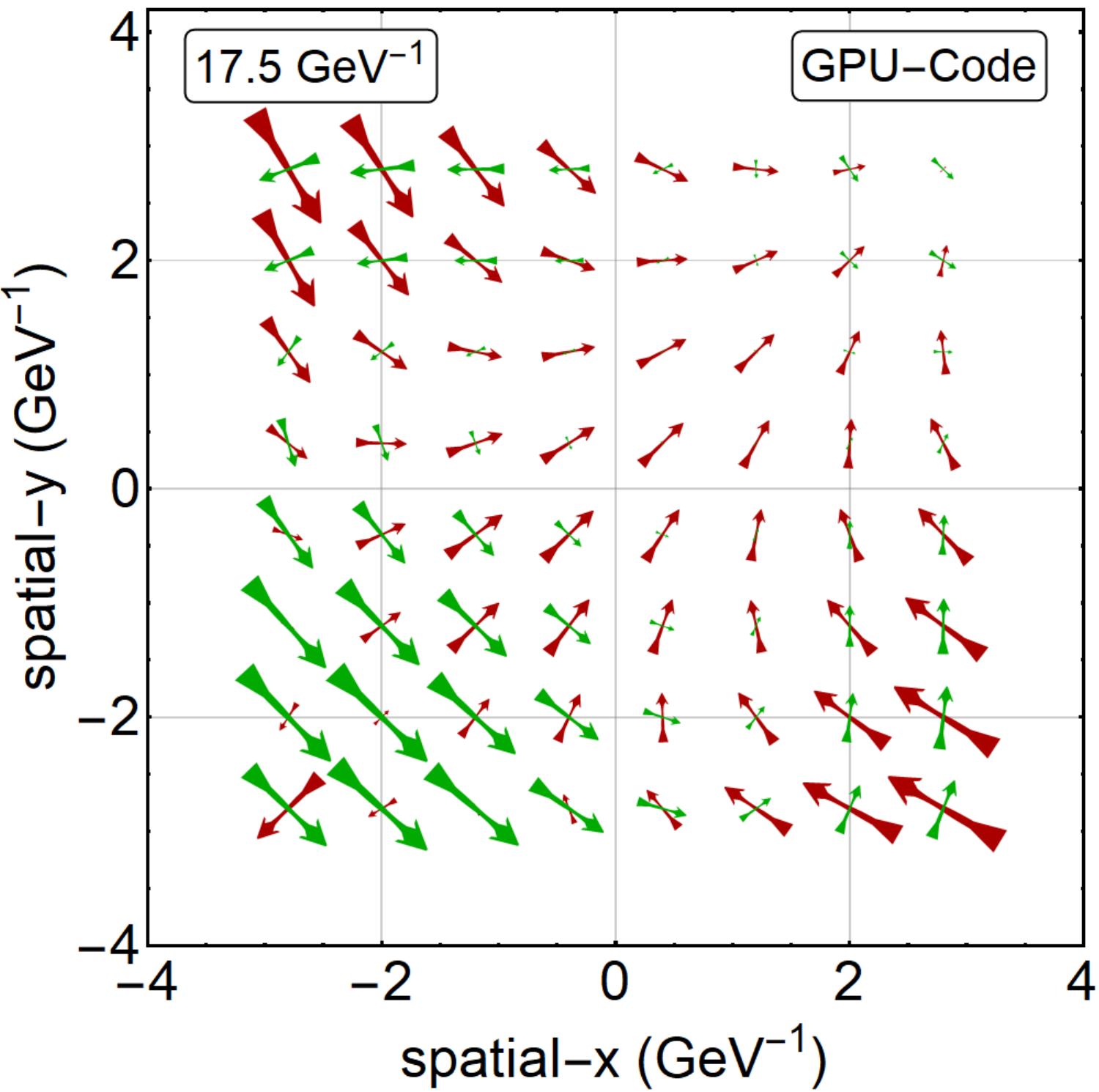}\tabularnewline
\end{tabular}
\par\end{centering}
\caption{Electric and magnetic field in the XOY plane with sinusoidal $\rho$
and $\mathbf{J}$. \label{fig:Electric-and-magnetic-1}}
\end{figure}

\begin{figure}
\begin{centering}
\begin{tabular}{ccc}
\includegraphics[scale=0.25]{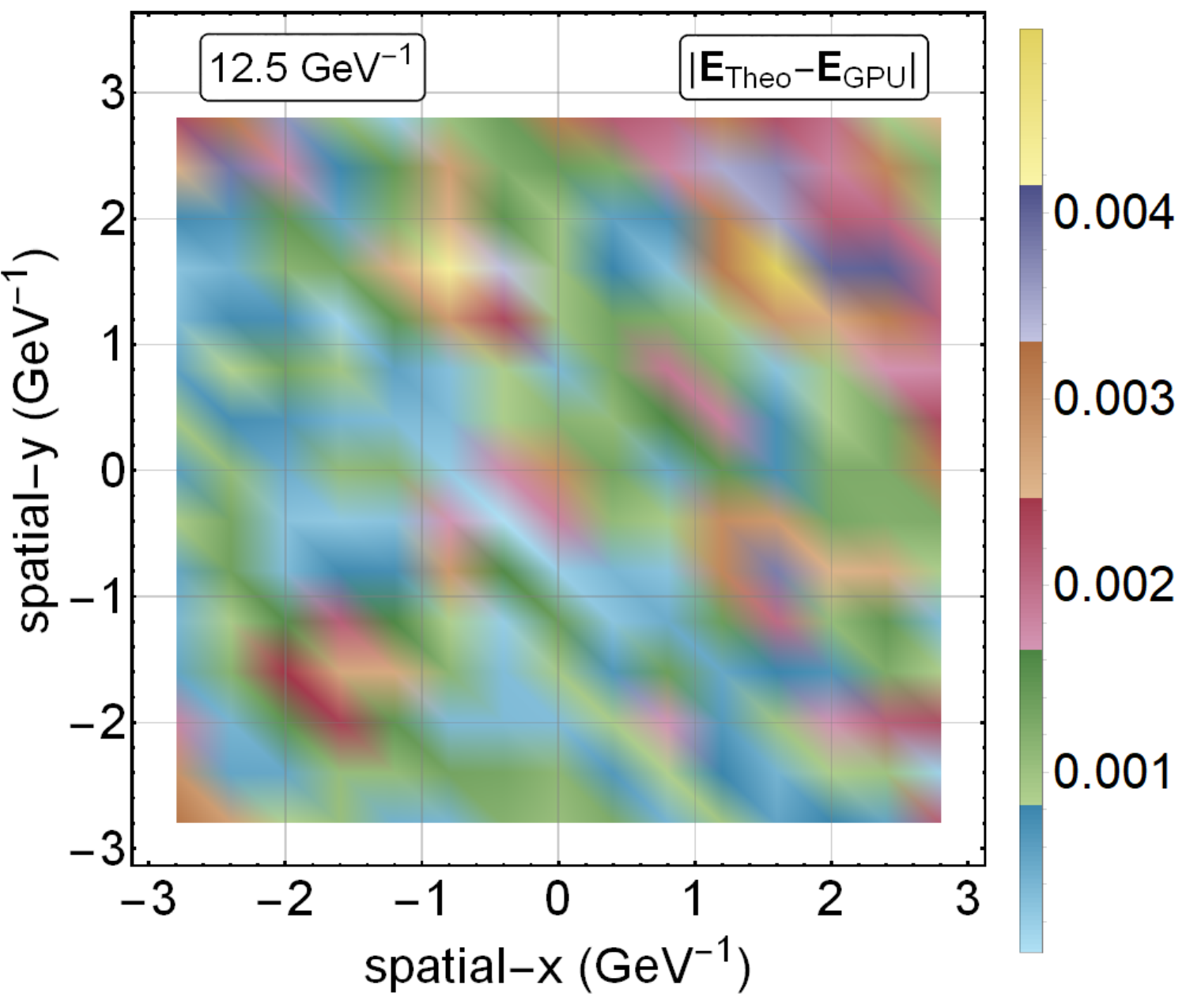} & \includegraphics[scale=0.25]{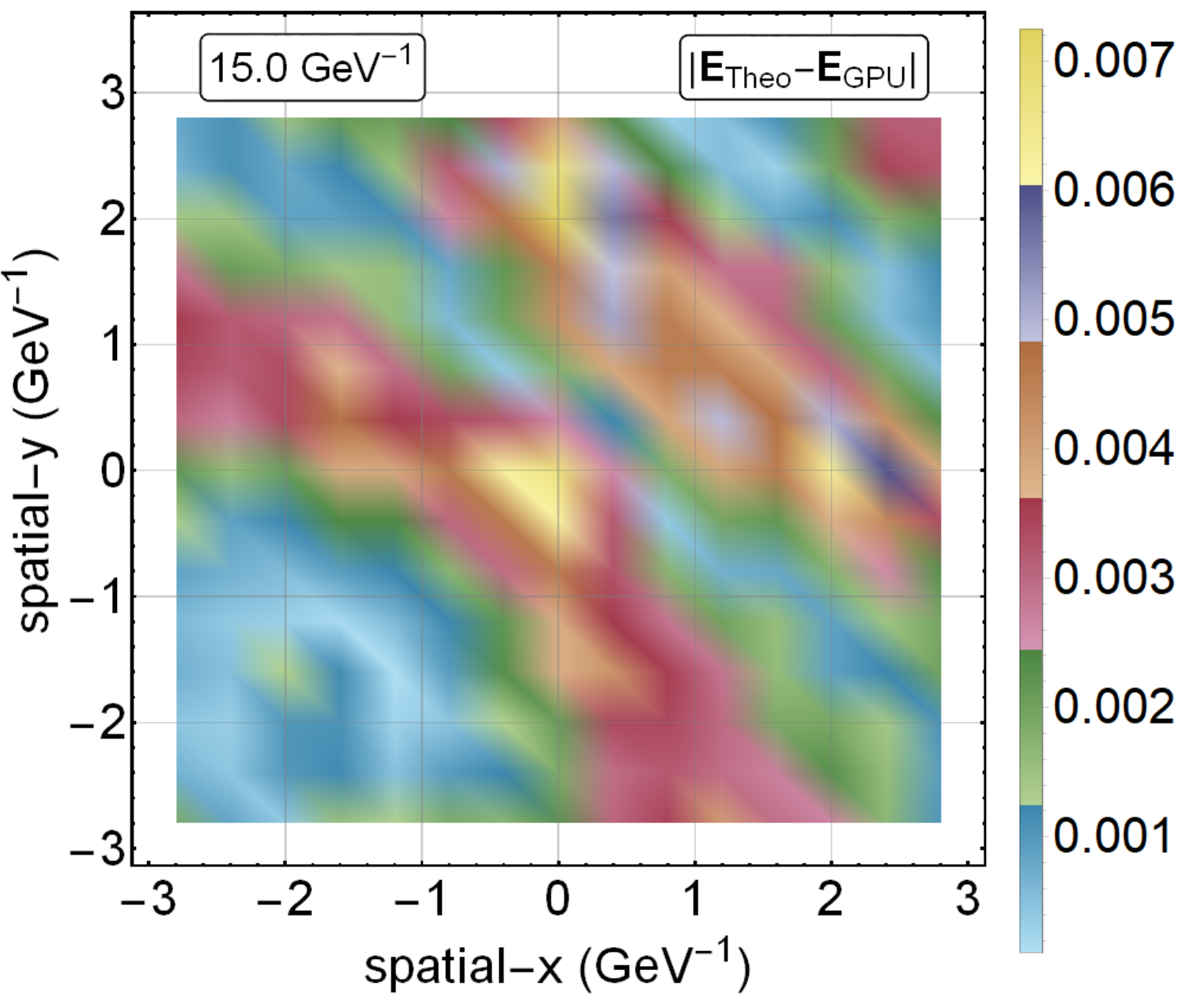} & \includegraphics[scale=0.25]{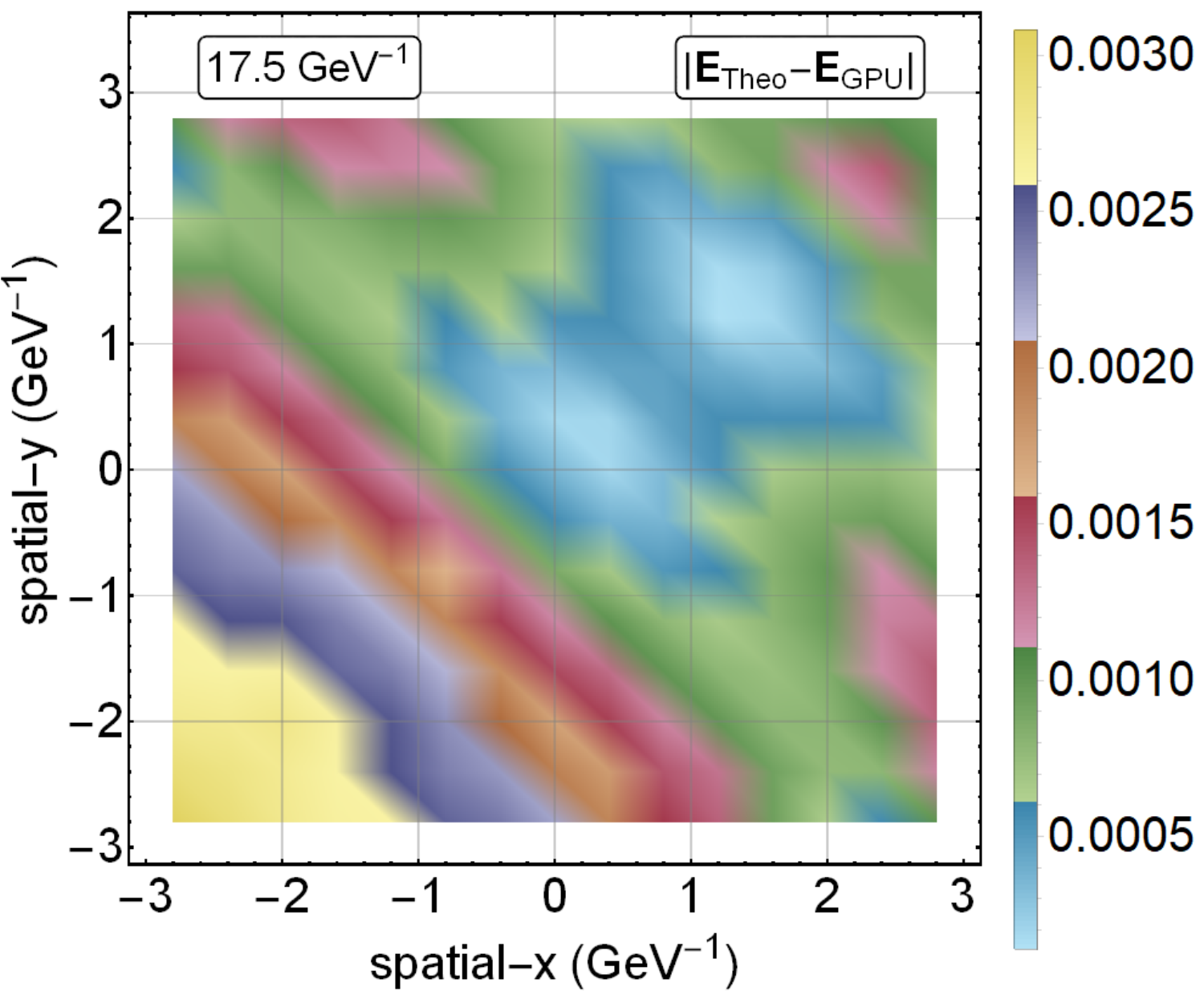}\tabularnewline
\includegraphics[scale=0.25]{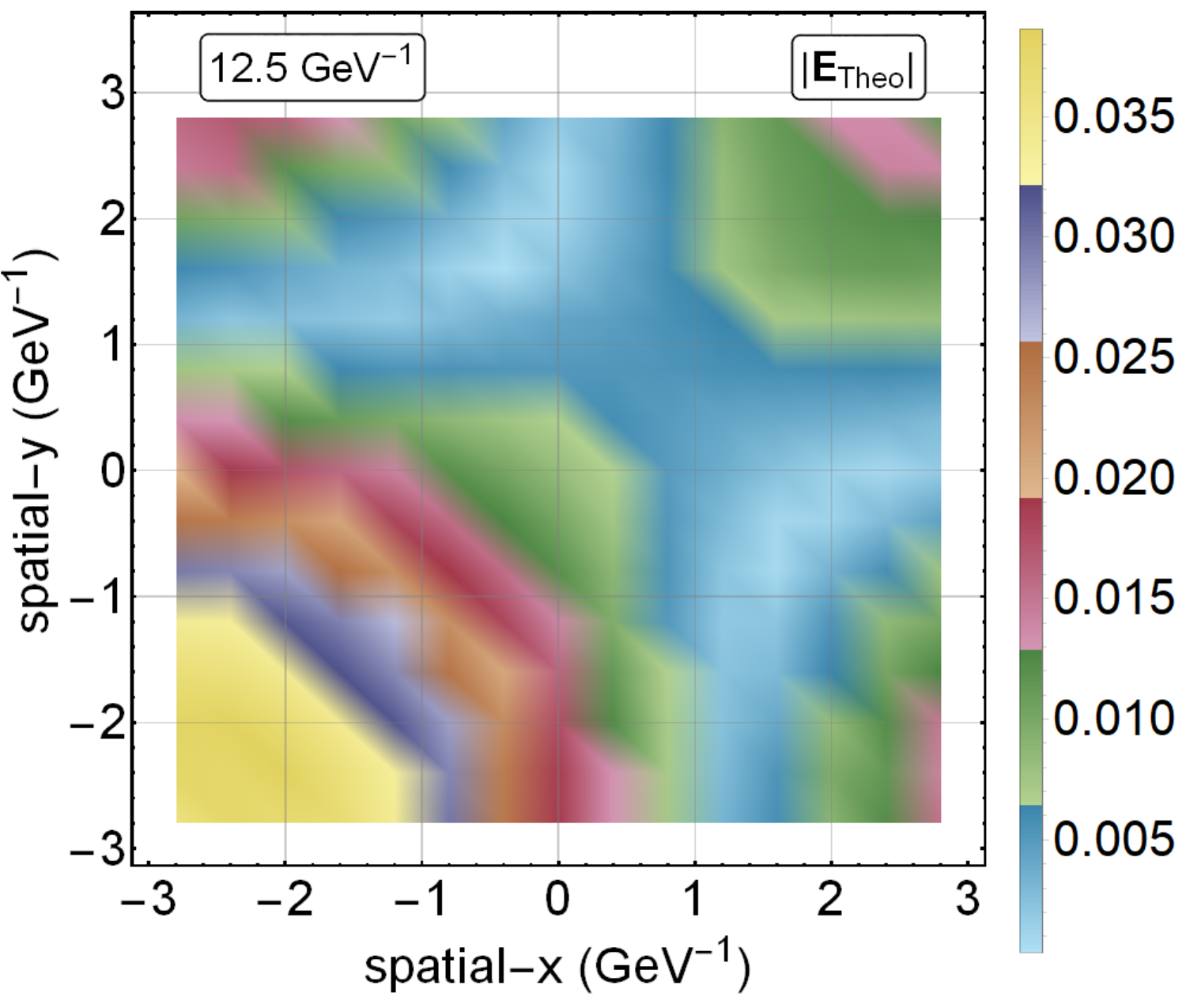} & \includegraphics[scale=0.25]{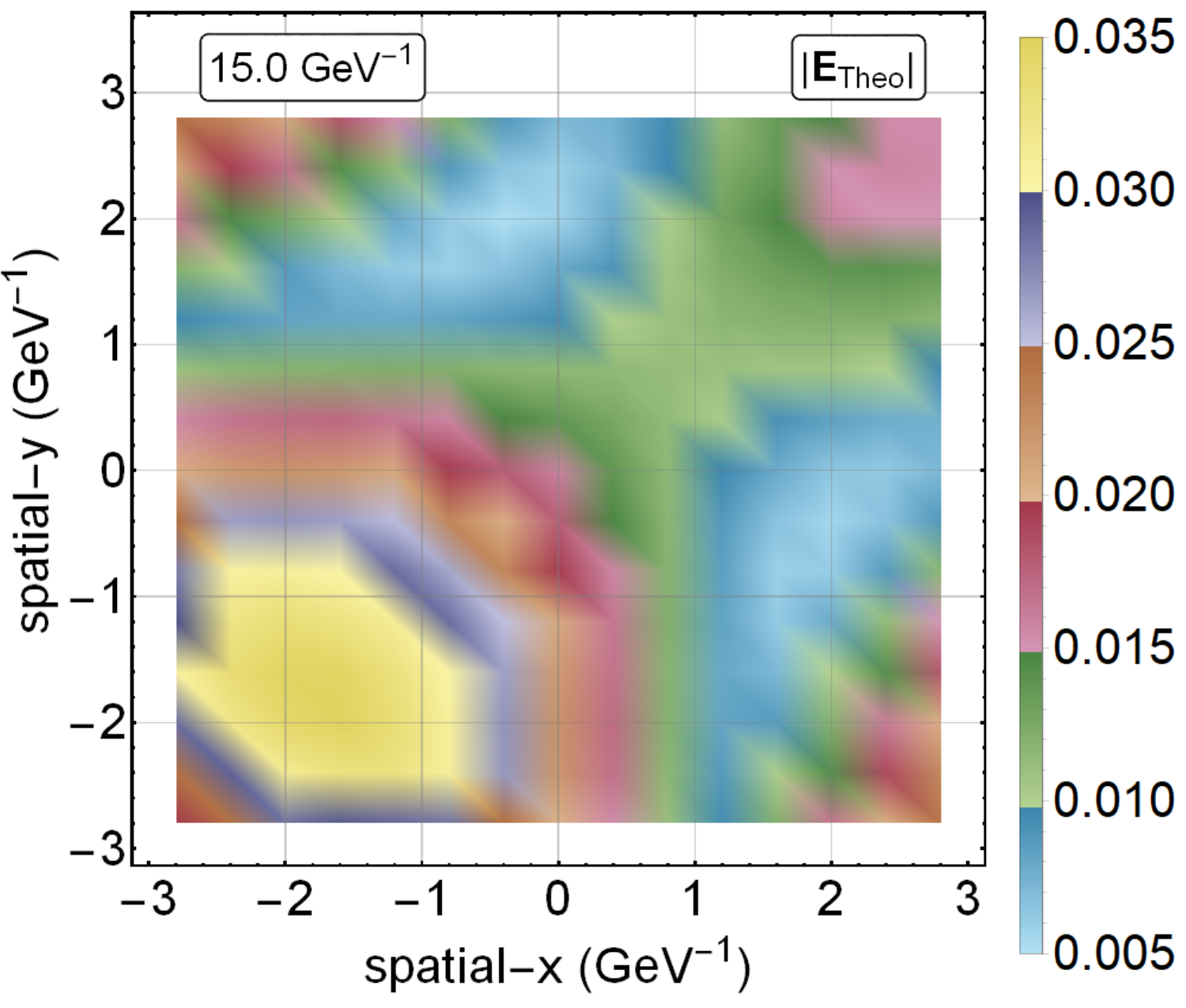} & \includegraphics[scale=0.25]{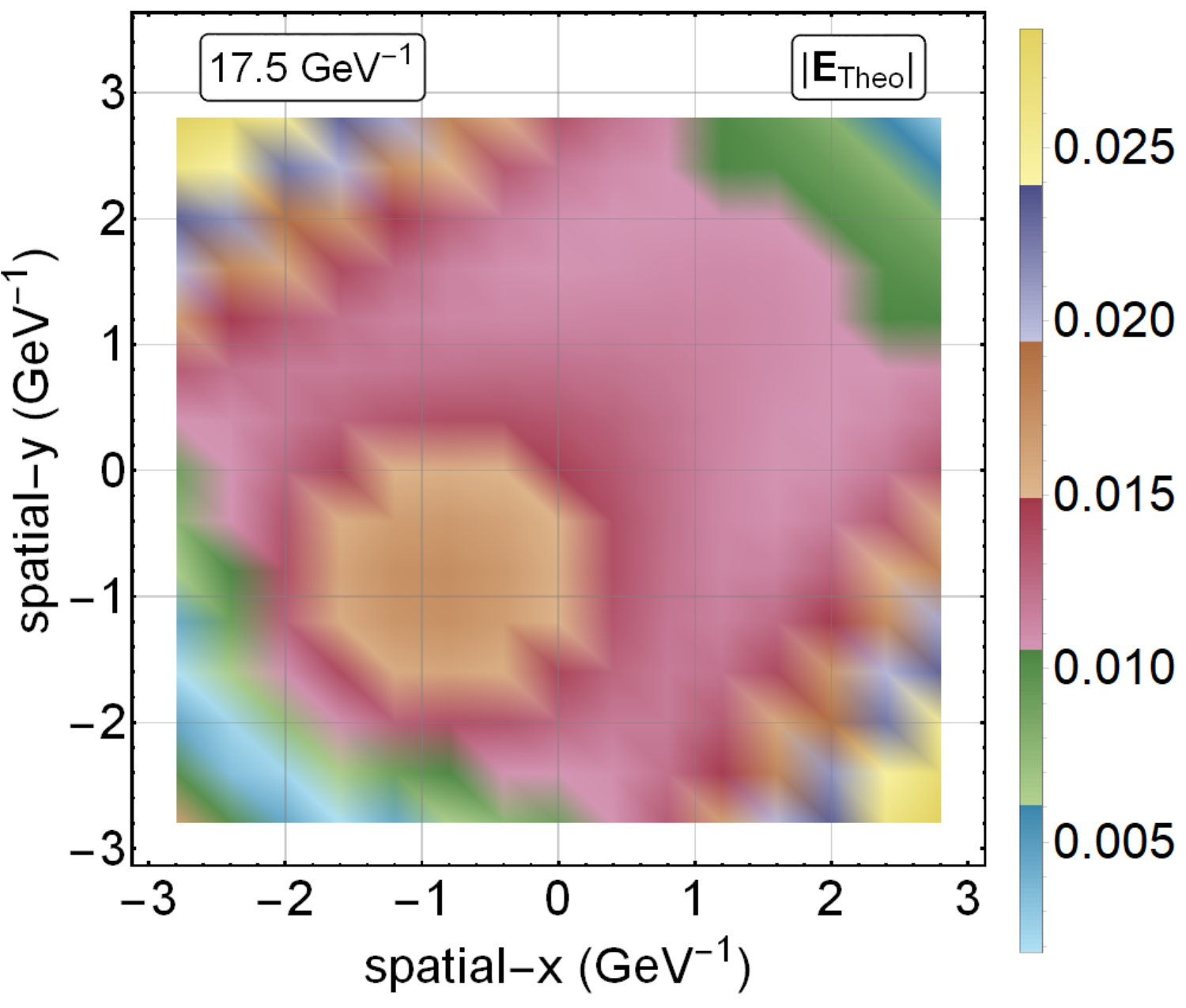}\tabularnewline
\end{tabular}
\par\end{centering}
\caption{Norm of $\mathbf{E}_{\text{Theo}}-\mathbf{E}_{\text{GPU}}$ and $\mathbf{E}_{\text{Theo}}$
with sinusoidal sources.\label{fig:Norm-of--3}}
\end{figure}

\begin{figure}
\begin{centering}
\begin{tabular}{ccc}
\includegraphics[scale=0.25]{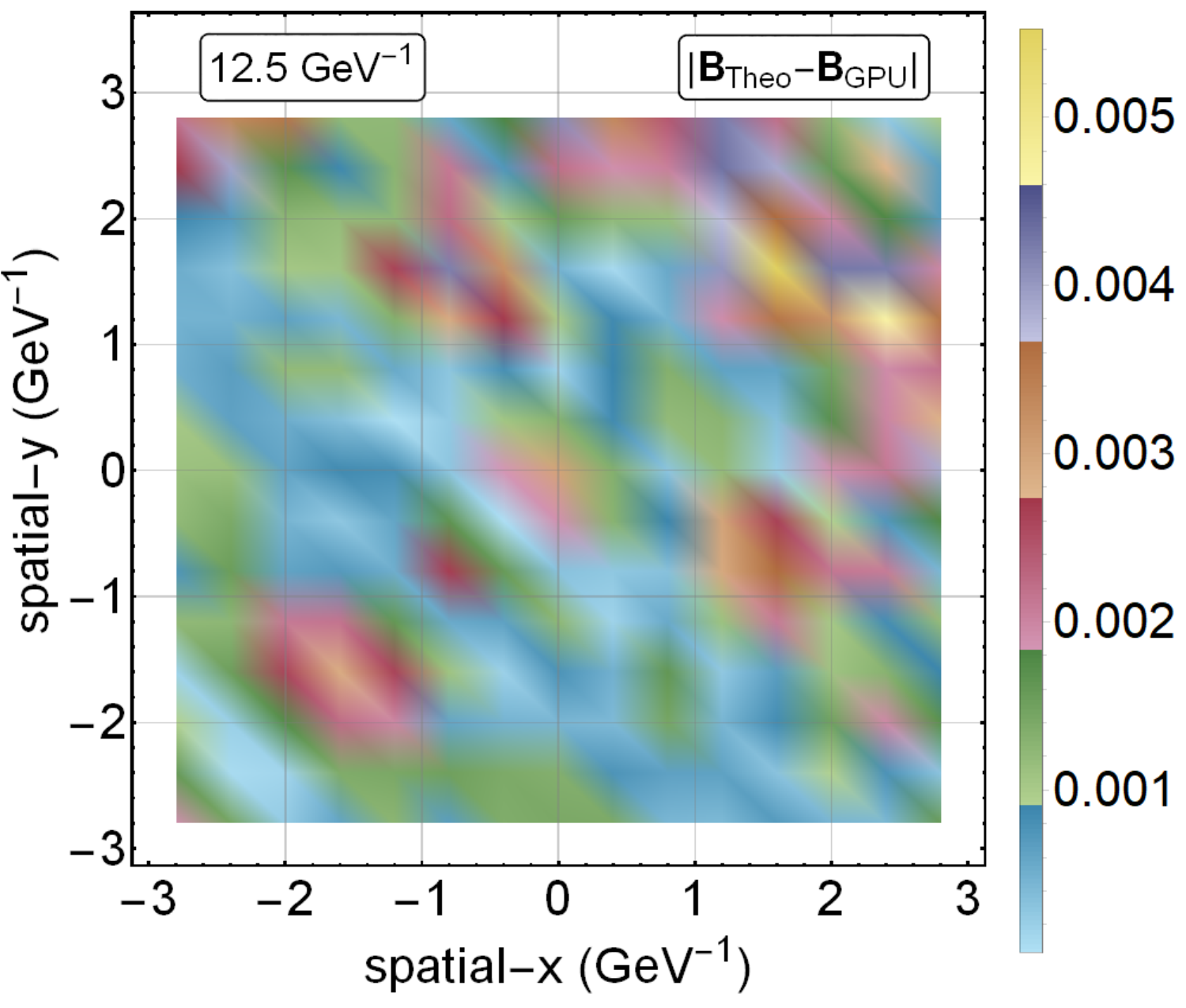} & \includegraphics[scale=0.25]{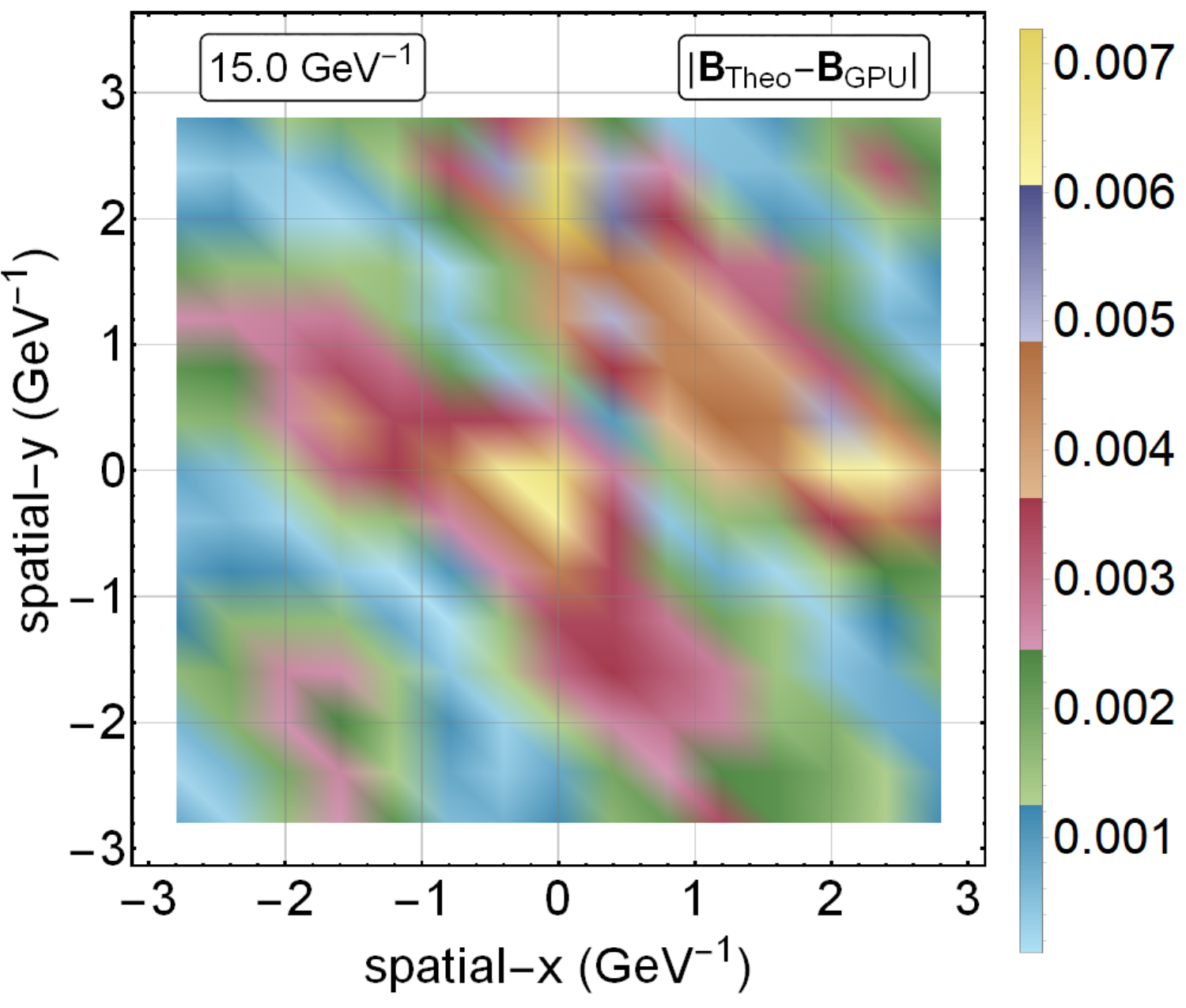} & \includegraphics[scale=0.25]{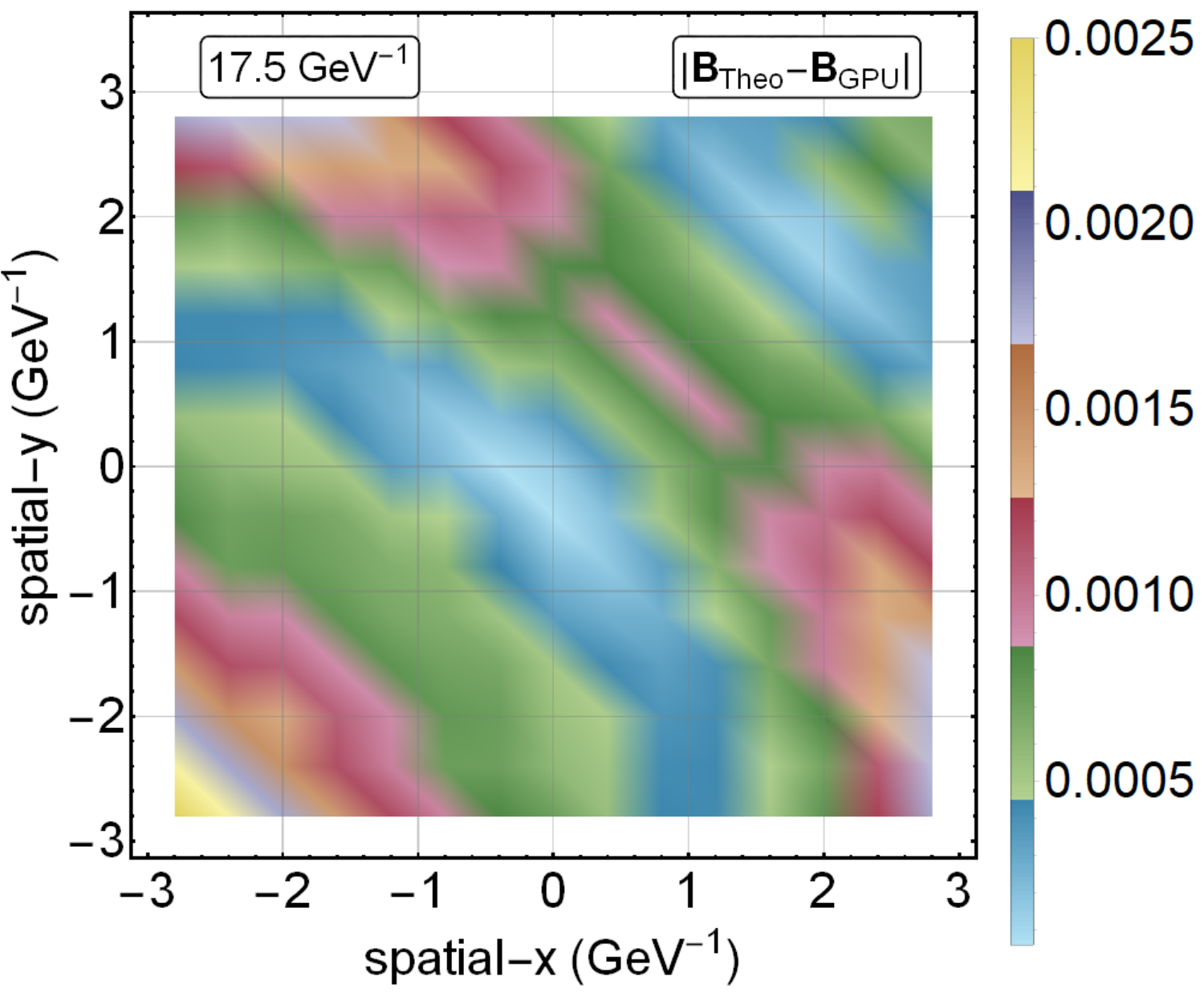}\tabularnewline
\includegraphics[scale=0.25]{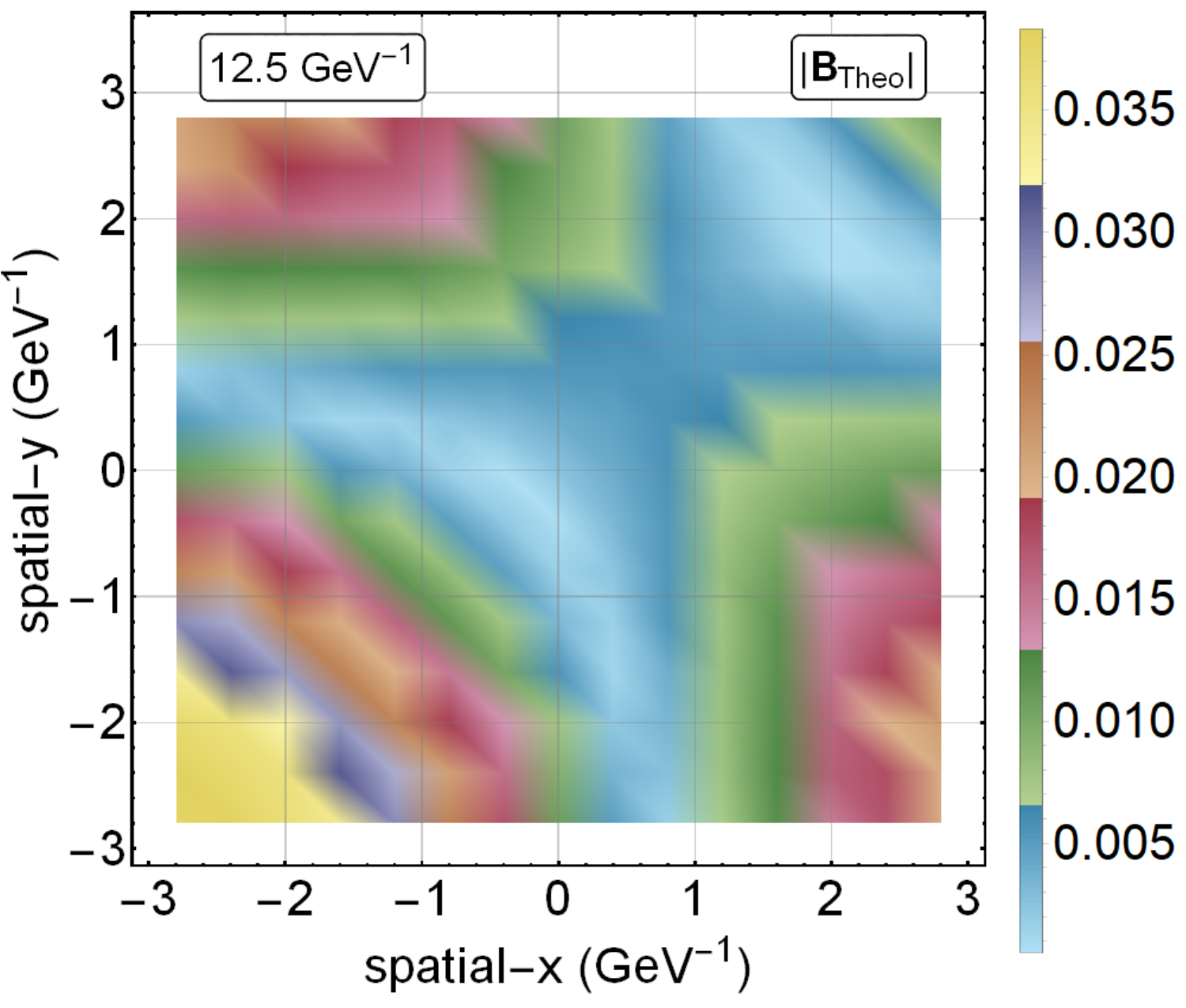} & \includegraphics[scale=0.25]{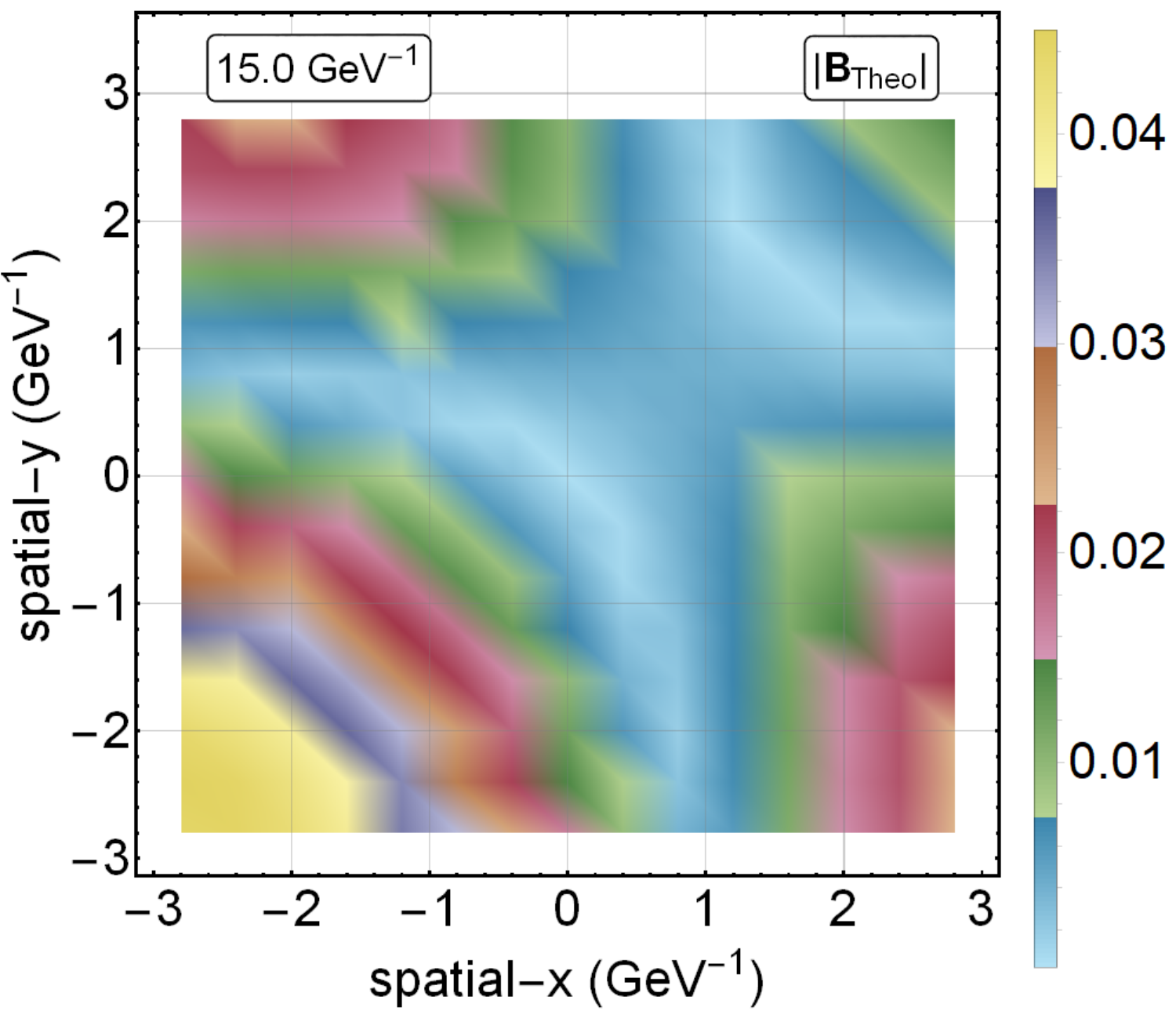} & \includegraphics[scale=0.25]{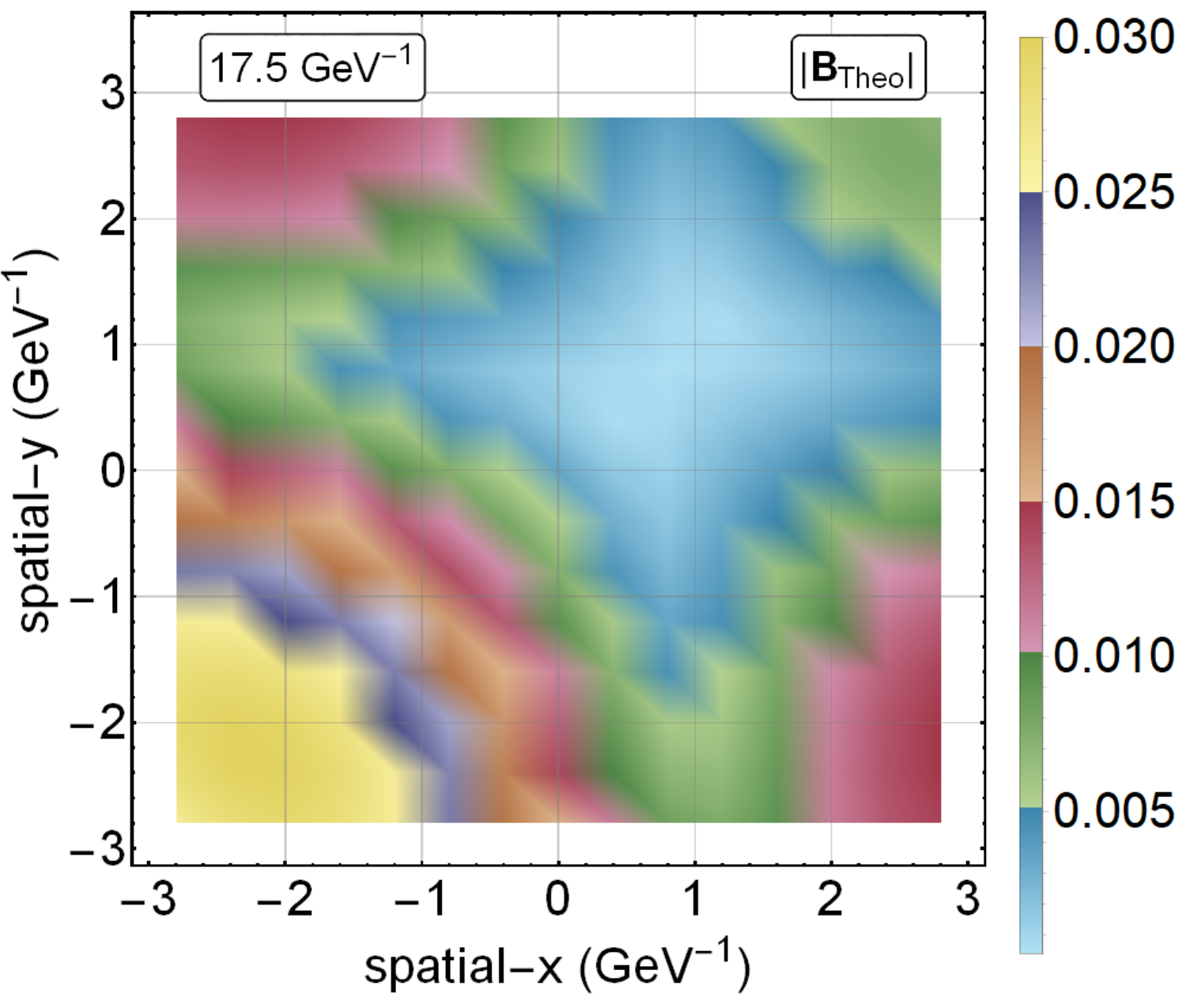}\tabularnewline
\end{tabular}
\par\end{centering}
\caption{Norm of $\mathbf{B}_{\text{Theo}}-\mathbf{B}_{\text{GPU}}$ and $\mathbf{B}_{\text{Theo}}$
with sinusoidal sources.\label{fig:Norm-of--2}}
\end{figure}

Fig. \ref{fig:Electric-and-magnetic-1} confirms that the patterns
of the EM fields obtained from the theoretical and the GPU code are
similar. Different from the case of the constant sources, there is
no saturation of the EM fields. Instead, we find the periodic features
of the EM fields which are consistent with the sinusoidal sources.
Fig. \ref{fig:Norm-of--3} and \ref{fig:Norm-of--2} show a maximum
deviation from the theoretical results of $0.004/0.035\sim11.2%\%
$ for $\mathbf{E}$ and $0.005/0.035\sim14.3%\%
$ for $\mathbf{B}$, respectively.

\subsection{Comparison of the execution time}

For a comparison of the execution time, we have performed a similar
algorithm depicted in Fig. \ref{fig:Schematic-of-the} on a CPU apparatus
whose hardware set-up is listed in Tab. \ref{tab:Hardware-set-up-of}.
We have chosen the same condition as in Sec. \ref{subsec:Constant-sources}.
Tab. \ref{tab:Comparison-of-the} shows the execution time of both
the CPU and GPU implementations. One can see that the use of GPU can
significantly enhance the performance by a factor of up to a thousand.
The detailed time consumption of HToD (Host to Device) and DToH (Device
to Host) will be demonstrated in Sec. \ref{sec:Parameter-dependence-of}.

\begin{table}
\caption{Comparison of the Execution time in GPU and CPU implementations. To
save time, the CPU code only evaluates the EM field for one time-step,
while the GPU code runs for totally 410 time snapshots ranging from
0 $\text{GeV}^{-1}$ to 20.5 $\text{GeV}^{-1}$. The two time-info
in the table correspond to the constant/sinusoidal sources (e.g.,
8.696/8.111 s means that 8.696 s is the execution time for the constant
source and 8.111 s for the sinusoidal source).The last row gives the
effective execution time where both implementations are assumed to
be executed for the same time snapshots.\label{tab:Comparison-of-the}}

\centering{}%
\begin{tabular}{|c|c|c|}
\hline 
Apparatus & GPU  & CPU\tabularnewline
\hline 
Execution time & 8.696/8.111 s & 18.012/18.540 s\tabularnewline
\hline 
Evaluated time snapshots & 410  & 1\tabularnewline
\hline 
Effective execution time & 8.696/8.111 s & 2.051/2.112 h\tabularnewline
\hline 
\end{tabular}
\end{table}

\section{Parameter dependence of the algorithm on one Tesla V100 card\label{sec:Parameter-dependence-of}}

There are two parameters that mainly affect the performance of the
algorithm, i.e., ``lts'' short for length\_of\_time\_snapshots and
``ntg'' short for number\_of\_total\_grids. In Eqs. (\ref{eq:E-1})
\textasciitilde{} (\ref{eq:tr-1}), the summations of the source terms
$\rho$ and $\mathbf{J}$ are at the retarded time $t_{r}$. Therefore
we need to store $\rho$ and $\mathbf{J}$ in the GPU memory at all
relevant times, i.e., $\rho_{\text{GPU}}\mapsto[\rho_{0}(\mathbf{r}),\rho_{1}(\mathbf{r}),\cdots\rho_{\text{lts}}(\mathbf{r})]$
and $\mathbf{J}_{\text{GPU}}\mapsto[\mathbf{J}_{0}(\mathbf{r}),\mathbf{J}_{1}(\mathbf{r}),\cdots\mathbf{J}_{\text{lts}}(\mathbf{r})]$.
If the maximum distance between the source point $\mathbf{r}_{\text{s}}$
and the observational point $\mathbf{r}_{\text{o}}$is $d_{\text{m}}=\text{max}\{|\mathbf{r}_{\text{s}}-\mathbf{r}_{\text{o}}|\}$,
and the time step used in the calculation is $dt$, then we can determine
the value of length\_of\_time\_snapshots by $\text{\textquotedblleft lts\textquotedblright}=\left\lceil d_{\text{m}}/dt\right\rceil $.
number\_of\_total\_grids counts the total grid sizes of both the observational
and source region, and is determined by the pair $\text{ \textquotedblleft ntg\textquotedblright}=[n_{x,\text{s}}\times n_{y,\text{s}}\times n_{z,\text{s}},n_{x,\text{o}}\times n_{y,\text{o}}\times n_{z,\text{o}}]$.
The total amount of float numbers on one GPU card can be evaluated
approximately as
\begin{eqnarray}
N & = & (\left\lceil d_{\text{m}}/dt\right\rceil +1)\times n_{x,\text{s}}\times n_{y,\text{s}}\times n_{z,\text{s}}\times4\nonumber \\
 &  & +n_{x,\text{o}}\times n_{y,\text{o}}\times n_{z,\text{o}}\times6,\label{eq:}
\end{eqnarray}
where number ``4'' corresponds to $\rho,J_{x},J_{y},J_{z}$ and
number ``6'' corresponds to $E_{x},E_{y},E_{z},B_{x},B_{y},B_{z}$.
We can use $N$ to evaluate the occupied GPU memory in the execution.
Apart from the memory consumption, the execution time is mainly constrained
by ``ntg'' since the summation and loop are on the spatial grids. 

\begin{table}

\caption{Occupied GPU memory at various ``ntg'' and ``lts''.\label{tab:Occupied-GPU-memory}}

\begin{centering}
\begin{tabular}{|cc|c|c|c|c|c|}
\hline 
 & ``ntg'' & $[160^{3},160^{3}]$ & $[100^{3},100^{3}]$ & $[70^{3},70^{3}]$ & $[60^{3},60^{3}]$ & $[20^{3},20^{3}]$\tabularnewline
``lts'' &  &  &  &  &  & \tabularnewline
\hline 
9000 &  &  &  &  & 30025MiB & 1461MiB\tabularnewline
\hline 
6000 &  &  &  & 31775MiB & 20137MiB & 1093MiB\tabularnewline
\hline 
2000 &  &  & 30899MiB & 10839MiB & 6953MiB & 605MiB\tabularnewline
\hline 
500 &  & 31707MiB & 8011MiB & 2991MiB & 2009MiB & 421MiB\tabularnewline
\hline 
\end{tabular}
\par\end{centering}
\end{table}
In Tab. \ref{tab:Occupied-GPU-memory}, we have given the GPU memory
consumption with different values of ``ntg'' and ``lts''. The
memory consumption is roughly consistent with Eq. \ref{eq:}. For
specific tasks, users can choose the proper parameter values according
to the GPU condition. Tab. \ref{tab:Time-consumption-at} shows the
time consumption of HToD, DToH and GPU execution for one time-step.
As a comparison with the CPU implementation, we have also conducted
a similar calculation with $\text{\textquotedblleft ntg\textquotedblright}=[20^{3},20^{3}]$
and $\text{\textquotedblleft lts\textquotedblright}=500$. The corresponding
execution time for one time-step in the CPU calculation is 102.25
s, which is around 100 times slower than the GPU implementation. However,
this does not mean that the GPU version is only 100 faster than the
CPU version. From Eqs. (\ref{eq:E-1}) \textasciitilde{} (\ref{eq:tr-1}),
we see that the execution time increases exponentially with \textquotedblleft ntg\textquotedblright .
Therefore, one should expect a significant performance improvement
with larger \textquotedblleft ntg\textquotedblright{} in the GPU implementation.
In our CPU version, we have only used one CPU core, therefore the
execution time should be extremely large for $\text{\textquotedblleft ntg\textquotedblright}\sim[70^{3},70^{3}]$,
and is impossible to perform.

\begin{table}
\caption{Time consumption with various values of ``ntg'' and ``lts''. The
upper and lower panels give the time consumption of {[}DToH, HToD{]}
and the GPU execution at each time step. \label{tab:Time-consumption-at}}

\begin{centering}
\begin{tabular}{|cc|c|c|c|c|}
\hline 
 & ``lts'' & 9000 & 6000 & 2000 & 500\tabularnewline
``ntg'' &  &  &  &  & \tabularnewline
\hline 
$[160^{3},160^{3}]$ &  &  &  &  & {[}0.060,0.042{]} s\tabularnewline
\hline 
$[100^{3},100^{3}]$ &  &  &  & {[}0.027,0.008{]} s & {[}0.027,0.009{]} s\tabularnewline
\hline 
$[70^{3},70^{3}]$ &  &  & {[}0.009,0.008{]} s & {[}0.009, 0.004{]} s & {[}0.009, 0.005{]} s\tabularnewline
\hline 
$[60^{3},60^{3}]$ &  & {[}0.005,0.001{]} s & {[}0.005,0.004{]} s & {[}0.006,0.003{]} s & {[}0.002,0.004{]} s\tabularnewline
\hline 
$[20^{3},20^{3}]$ &  & {[}0.0004,0.016{]} s & {[}0.0004,0.002{]} s & {[}0.0004,0.002{]} s & {[}0.0004,0.002{]} s\tabularnewline
\hline 
\end{tabular}
\par\end{centering}
\centering{}%
\begin{tabular}{|cc|c|c|c|c|}
\hline 
 & ``lts'' & 9000 & 6000 & 2000 & 500\tabularnewline
``ntg'' &  &  &  &  & \tabularnewline
\hline 
$[160^{3},160^{3}]$ &  &  &  &  & 521.970 s\tabularnewline
\hline 
$[100^{3},100^{3}]$ &  &  &  & 32.583 s & 32.565 s\tabularnewline
\hline 
$[70^{3},70^{3}]$ &  &  & 4.845 s & 4.813 s & 4.858 s\tabularnewline
\hline 
$[60^{3},60^{3}]$ &  & 2.706 s & 2.668 s & 2.679 s & 2.720 s\tabularnewline
\hline 
$[20^{3},20^{3}]$ &  & 1.006 s & 1.008 s & 1.067 s & 1.001 s\tabularnewline
\hline 
\end{tabular}
\end{table}

\section{Conclusion\label{sec:Conclusion}}

In the current work, we have provided an implementation of the Jefimenko's
equations on GPU based on our previous works\cite{Zhang2020,Wu2019tsf,zhang2020tsf}.
Our code gives stable and accurate results compared with the theoretical
calculations. To see the enhancement of the GPU code, we have also
performed a similar algorithm in with 1 CPU core available. A significant
improvement of the execution time has been observed in the GPU implementation.
Finally, we give the parameter dependence of the GPU code on one NVIDIA
Tesla V100 card. The simulation results in Tab. \ref{tab:Time-consumption-at}
shows that the execution time of the GPU implementation is at least
100 times faster compared with single CPU. 

In the API we have provided users with scaling functionalities. With
specified domains of both the observational and source regions, the
code can be easily used in multi-GPU clusters. However, the current
version of the code only supports multi-GPU manipulations via the
Ray package. For an automatic use of these scaling, we still require
a heuristic module to allocate the GPUs. Meanwhile, the current version
does not support expanding systems like the condition in heavy-ion
collisions. We will consider all these factors in the future.

\label{}



%% References


\bibliographystyle{elsarticle-num}
\bibliography{Ref}






% \end{thebibliography}
\end{document}